\documentclass[11pt,a4paper]{article}
\pdfoutput=1 

\usepackage{bm}
\usepackage{calrsfs}
\usepackage{caption}
\usepackage{xcolor}
\usepackage{jheppub} 
\usepackage{siunitx}
\usepackage{cleveref}

\usepackage[T1]{fontenc} 

\crefname{section}{section}{sections}
\Crefname{section}{Section}{Sections}

\crefname{figure}{figure}{figures}
\Crefname{figure}{Figure}{Figures}
\crefrangeformat{figure}{figures~#3#1#4--#5#2#6}
\Crefrangeformat{figure}{Figures~#3#1#4--#5#2#6}

\crefrangeformat{equation}{eqs.~(#3#1#4)--(#5#2#6)}
\Crefrangeformat{equation}{Equations~(#3#1#4)--(#5#2#6)}

\creflabelformat{subequations}{(#2#1#3)}
\crefname{subequations}{eqs.}{eqs.}
\crefrangeformat{subequations}{eqs.~(#3#1#4)--(#5#2#6)}
\Crefname{subequations}{Equations}{Equations}
\Crefrangeformat{subequations}{Equations~(#3#1#4)--(#5#2#6)}

\newcommand{\as}{\ensuremath{\alpha_\text{s}}}
\newcommand{\abar}{\ensuremath{\alpha_{\overline{\rm MS}}}}
\newcommand{\ag}{\ensuremath{\alpha_\text{g}}}
\newcommand{\an}[1]{\ensuremath{\alpha_{#1}}}
\newcommand{\dd}{\ensuremath{\mathrm{d}}}
\newcommand{\e}{\ensuremath{\mathrm{e}}}
\newcommand{\iunit}{\ensuremath{\mathrm{i}}}
\newcommand{\case}[2]{\ensuremath{{\textstyle\frac{#1}{#2}}}}
\newcommand{\half}{\ensuremath{\case{1}{2}}}
\newcommand{\third}{\ensuremath{\case{1}{3}}}
\newcommand{\sixth}{\ensuremath{\case{1}{6}}}

\newcommand{\MSbar}{\ensuremath{\overline{\rm MS}}}

\newlength{\figwidth}

\allowdisplaybreaks

\subheader{\sf FERMILAB-PUB-23-629-T}
   \title{Factorial growth at low orders in perturbative~QCD: Control over truncation uncertainties}

\author{Andreas S. Kronfeld}

\affiliation{Theory Division, Fermi National Accelerator Laboratory,\\
    P.O. Box 500, Batavia IL 60565, USA}
\affiliation{Institute for Advanced Study, Technische Universität München,\\
    Lichtenbergstraße 2a, 85748 Garching bei München, Germany}

\emailAdd{ask@fnal.gov} 

\abstract{A method, known as ``minimal renormalon subtraction''
[\href{https://inspirehep.net/literature/1643281}{\emph{Phys.~Rev.~D} \textbf{97} (2018) 034503},
\href{https://inspirehep.net/literature/1507484}{\emph{JHEP} \textbf{2017} (2017) 62}], relates the factorial growth of a
perturbative series (in QCD) to the power~$p$ of a power correction $\Lambda^p/Q^p$.
($\Lambda$ is the QCD scale, $Q$ some hard scale.) %
Here, the derivation is simplified and generalized to any~$p$, more than one such correction, and cases with anomalous dimensions.
Strikingly, the well-known factorial growth is seen to emerge already at low or medium orders, as a consequence of constraints on
the $Q$ dependence from the renormalization group.
The effectiveness of the method is studied with the gluonic energy between a static quark and static antiquark (the ``static
energy'').
Truncation uncertainties are found to be under control after next-to-leading order, despite the small exponent of the power
correction ($p=1$) and associated rapid growth seen in the first four coefficients of the perturbative series.}

\arxivnumber{2310.15137}

\begin{document}
\maketitle
\flushbottom

\section{Introduction}
\label{sec:intro}

In 2018, the Fermilab Lattice, MILC, and TUMQCD collaborations~\cite{FermilabLattice:2018est} used lattice-QCD calculations of
heavy-light meson masses to obtain results for renormalized quark masses in the modified minimal subtraction (\MSbar) scheme.
The total uncertainty ranges from below 1\% (for bottom, charm, and strange) to 1--2\% (for down and up).
The \MSbar\ scheme inevitably entails perturbation theory.
Usually a top source of uncertainty would come from truncating the perturbative series in the strong coupling \as.
In ref.~\cite{FermilabLattice:2018est}, however, the error budgets exhibit negligible uncertainty from truncation (cf.,
figure~4~\cite{FermilabLattice:2018est}).
The associated uncertainty was estimated by omitting the highest-order coefficient (of $\as^4$) in the relation between the pole
mass and the \MSbar\ mass.
It was found to be comparable to the statistical uncertainty and much smaller than the parametric uncertainty in \as.

Essential to ref.~\cite{FermilabLattice:2018est} is a reinterpretation of the perturbation series~\cite{Brambilla:2017hcq} that in
turn relies crucially on a formula for the normalization of the leading renormalon ambiguity of the pole
mass~\cite{Komijani:2017vep}.
Readers who are not familiar with renormalons are encouraged to indulge the jargon for a moment: clearly it is worth pursuing how to
generalize refs.~\cite{Komijani:2017vep,Brambilla:2017hcq}, in the hope of controlling the truncation uncertainty in further
applications.
This paper takes up that pursuit.

The coefficients of many perturbative series in quantum mechanics~\cite{Bender:1971gu,Bender:1973rz} and quantum field
theory~\cite{Gross:1974jv,Lautrup:1977hs,tHooft:1977xjm} are known to grow factorially.
In QCD and other asymptotically free theories, a~class of leading and subleading growths arises from soft loop momenta in Feynman
diagrams.
Details of the growth can be obtained from studying implications of the renormalization group.
At the same time, the growth is related to power-law corrections to the perturbation series.
For now, let us characterize the growth of the $l$\textsuperscript{th} coefficient as $Ka^l l^b l!$ for some $K$, $a$, and~$b$.
A basic renormalization-group analysis (e.g., ref.~\cite{Beneke:1998ui}) determines $a$ and $b$ but not the normalization~$K$.
There are, however, at least three expressions in the literature for
$K$~\cite{Komijani:2017vep,Lee:1996yk,Lee:1999ws,Pineda:2001zq,Hoang:2008yj}.
The expressions in refs.~\cite{Komijani:2017vep} and~\cite{Hoang:2008yj} bear some resemblance to each other, but the one in
refs.~\cite{Lee:1996yk,Lee:1999ws,Pineda:2001zq} is different.

The generalizations initially sought in the present work started modest: I wanted to look at scale dependence of \as\ to see (as a
co-author of refs.~\cite{FermilabLattice:2018est,Brambilla:2017hcq}) whether our quoted uncertainties held up, and I wanted to treat
arbitrary power corrections.
Dissatisfaction with my understanding of the normalization derived in ref.~\cite{Komijani:2017vep} led to a simple way of analyzing
the problem with interesting findings:
\begin{itemize}
    \item the normalization of ref.~\cite{Komijani:2017vep} is reproduced, at least in practical terms;
    \item the standard factorial growth starts at low orders, not just at asymptotically large~$l$;
    \item the second coefficient of the $\beta$ function and the exponent of the power correction determine the order at which the
        factorial growth becomes a practical matter; \item the way to deal with a sequence of power corrections becomes clear.
\end{itemize}
The third item is well known, but, even so, many analyses of large-order effects use a one-term $\beta$ function.
The last item was mentioned in v1 and v2 on \href{https://arXiv.org/abs/1701.00347v2}{arXiv.org} of ref.~\cite{Komijani:2017vep},
but the discussion was removed from the final publication.
The derivation of the factorial growth presented below is so straightforward, it is almost surprising that it has not been known for
decades.
If it has appeared in the literature before, it is obscure.

The rest of this paper is organized as follows.
\Cref{sec:MRS} recalls ref.~\cite{Brambilla:2017hcq} and generalizes its ideas to an arbitrary (single) power correction.
\Cref{sec:FRS} considers cases with more than one power-suppressed contribution.
\Cref{sec:MRS,sec:FRS} rely on a special renormalization scheme that simplifies the algebra; other schemes are discussed in
\cref{sec:scheme}.
\Cref{sec:anom} considers the complication of anomalous dimensions.
Proposals to improve perturbation theory should study at least one example, so \cref{sec:E0} applies \cref{sec:MRS} to the static
energy between a heavy quark-antiquark pair, for which four terms in the perturbation series are known (like the
pole-mass--\MSbar-mass relation).
A~summary and some outlook is offered in \cref{sec:outlook}.
A~modification of the Borel summation used in \cref{sec:MRS,sec:FRS} is given in \cref{app:MBS}.

\section{Notation and setup}
\label{sec:setup}

The problem at hand is to compute in QCD, or other asymptotically free quantum field theory, a physical quantity that depends on a
high-energy scale $Q$ (or, as in \cref{sec:E0}, short distance $r=1/Q$).
The hard scale $Q$ can be used to obtain a dimensionless version of the physical quantity.
\pagebreak
The dimensionless quantity can be approximated order-by-order in perturbation theory up to power corrections:
\begin{equation}
    \mathcal{R}(Q) = r_{-1} + R(Q) + C_p\frac{\Lambda^p}{Q^p}, \qquad
             R (Q) = \sum_{l=0} r_l(\mu/Q) \as(\mu)^{l+1},
    \label{eq:R}
\end{equation}
where the term $r_{-1}$ can be 0 or not, $C_p$ is (for now) independent of~$Q$, $\Lambda\sim\mu\e^{-1/2\beta_0\as(\mu)}$ is the
scale arising from dimensional transmutation, $\as$ is the gauge coupling in some scheme, and $\mu$ is the renormalization scale.
The power~$p$ can be deduced from the operator-product expansion, an effective field theory, or other considerations.
For now, let us consider the case with only one power correction, postponing until \cref{sec:FRS} the more general case.
Laboratory measurements or the continuum limit of lattice gauge theory can be used to provide a nonperturbative determination
of~$\mathcal{R}(Q)$.
Fits of data for~$\mathcal{R}(Q)$ could, ideally, be used to determine \as\ with nuisance parameter $C_p$.
As an asymptotic expansion, the sum representing $R(Q)$ in \cref{eq:R} diverges, however, so an upper summation limit does not make
sense without further discussion.
Indeed, the definition of the power correction rests on how the sum is treated.

$\mathcal{R}$ and $R$ do not depend of~$\mu$, so the $\mu$ dependence of the coefficients is intertwined with the $\mu$ dependence
of \as\ and, thus, dictated by
\begin{equation}
    \dot{\alpha}_\text{s}(\mu) \equiv 2\beta(\as) = - 2 \as(\mu) \sum_{k=0}^\infty \beta_k \as(\mu)^{k+1} ,
    \label{eq:beta}
\end{equation}
where $\dot{g}=\dd g/\dd\ln\mu$.
The derivatives of the coefficients must satisfy
\begin{equation}
    \dot{r}_l(\mu/Q) = 2\sum_{j=0}^{l-1} (j+1) \beta_{l-1-j} r_j(\mu/Q) .
    \label{eq:dot-r}
\end{equation}
Integrating these equations (in a mass-independent renormalization scheme) one after the other leads to
\begin{subequations} \label[subequations]{eq:r}
\begin{align}
    r_0(\mu/Q) = r_0 &,                                                                                    \label{eq:r0s} \\
    r_1(\mu/Q) = r_1 &+ 2 \beta_0\ln(\mu/Q) r_0 ,                                                          \label{eq:r1s} \\
    r_2(\mu/Q) = r_2 &+ 2\ln(\mu/Q)\left(2 \beta_0 r_1 + \beta_1 r_0\right) + [2\beta_0\ln(\mu/Q)]^2 r_0 , \label{eq:r2s} \\
    r_3(\mu/Q) = r_3 &+ 2\ln(\mu/Q)\left(3 \beta_0 r_2 + 2 \beta_1 r_1 + \beta_2 r_0\right) +
        3 [2\beta_0\ln(\mu/Q)]^2 r_1 \nonumber \\
        &+ 10 \beta_0\beta_1 \ln^2(\mu/Q) r_0 + [2\beta_0\ln(\mu/Q)]^3 r_0 ,                               \label{eq:r3s}
\end{align}
\end{subequations}
and so on, with constants of integration $r_l\equiv r_l(1)$.
The dependence of $R(Q)$ on $Q$ is, thus, tied to the renormalization-dictated dependence on~$\mu$.

\Cref{eq:dot-r} is a matrix equation, $\dot{\bm{r}}=2\mathbf{D}\cdot\bm{r}$, with $D_{lj}=(j+1)\beta_{l-1-j}$ if $l>j$ and
$D_{lj}=0$ otherwise.
For \cref{sec:MRS,sec:FRS,sec:scheme}, it is convenient to develop this matrix notation further, for instance writing
\begin{equation}
    R = \boldsymbol{\mathfrak{A}}_\text{s}\cdot\bm{r}_\text{s} =
        \left\lceil \as \quad \as^2 \quad \as^3 \quad \as^4 \quad \cdots \; \right\rceil
        \left\lceil \begin{array}{c}
            r_0 \\ r_1  \\ r_2 \\ r_3 \\ \vdots
        \end{array}
        \right\rceil .
\end{equation}
Floorless delimiters $\lceil\;\rceil$ are used instead of brackets $[\;]$ or parentheses as a reminder that the vectors are infinite
sequences.
Below it will be useful to think of the subscript ``$\text{s}$'' as standing for ``starting scheme'', in practice~\MSbar.

The matrix notation makes scheme and scale dependence manifest and eases derivations.
For example, if
\begin{equation}
    \alpha_b = \as + b_1 \as^2 + b_2 \as^3 + b_3 \as^4 + \cdots,
    \label{eq:ag:as}
\end{equation}
then $\boldsymbol{\mathfrak{A}}_b = \boldsymbol{\mathfrak{A}}_\text{s}\cdot\mathbf{b}^{-1}$ with scheme-conversion matrix
\begin{equation}
    \mathbf{b}^{-1} = \left\lceil\begin{array}{ccccc}
      1 \, &      0     &    0   &  \,0\, & \cdots \\
     b_1\, &      1     &    0   &  \,0\, & \cdots \\
     b_2\, &    2b_1    &    1   &  \,0\, & \cdots \\
     b_3\, & b_1^2+2b_2 &  3b_1  &  \,1\, & \cdots \\
    \vdots & \vdots     & \vdots & \ddots & \ddots \\
    \end{array}\right\rceil ,
    \hspace*{16pt}
    \mathbf{b}      = \left\lceil\begin{array}{ccccc}
             1         &       0     &    0   &  \,0\, & \cdots \\
           -b_1        &       1     &    0   &  \,0\, & \cdots \\
        2b_1^2-b_2     &    -2b_1    &    1   &  \,0\, & \cdots \\
    5b_1b_2-5b_1^3-b_3 & 5b_1^2-2b_2 & -3b_1  &  \,1\, & \cdots \\
    \vdots             & \vdots      & \vdots & \ddots & \ddots \\
    \end{array}\right\rceil .
    \label{eq:conv-b}
\end{equation}
The coefficients in the ``$b$'' scheme are $\bm{r}_b=\mathbf{b}\cdot\bm{r}_\text{s}$.
The lower-triangular structure of these and other matrices is the key to the forthcoming analysis.

The \MSbar\ scheme can be thought of as the ``laboratory frame'', where $\bm{r}_\text{s}$ is most easily obtained.
The ``center-of-mass frame'', which reduces subsequent labor, is the ``geometric scheme'' defined by~\cite{Brown:1992pk}
\begin{equation}
    \beta(\ag) = -\frac{\beta_0\ag^2}{1-(\beta_1/\beta_0)\ag}.
    \label{eq:betag}
\end{equation}
Equivalently, $\beta_k=\beta_0(\beta_1/\beta_0)^k$, so the $\beta$-function series, \cref{eq:beta}, is geometric.
In \cref{eq:ag:as}, $b_1=2\beta_0\ln\Lambda_\text{g}/\Lambda_{\MSbar}$; taking $b_1=0$ not only eliminates or simplifies many
entries in the scheme-conversion matrix but also means $\Lambda_\text{g}=\Lambda_{\MSbar}$ requires no conversion.
Expressions for the $b_i$ connecting the geometric and \MSbar\ schemes are less interesting than the entries of the conversion
matrix:
\begin{equation}
    \newcommand{\cb}{\third\delta_4 - \sixth\delta_3\check{\beta}_1 + \frac{5}{3}\delta_2^2 + \third\delta_2\check{\beta}_1^2}
    \mathbf{b}_\text{g} = \left\lceil\begin{array}{cccccc}
             1    &       0       &       0       &  \,0\, &  \,0\, & \cdots \\
             0    &       1       &       0       &  \,0\, &  \,0\, & \cdots \\
        \delta_2  &       0       &       1       &  \,0\, &  \,0\, & \cdots \\
    \half\delta_3 & \;2\delta_2\; &       0       &  \,1\, &  \,0\, & \cdots \\
           \cb    &    \delta_3   & \;3\delta_2\; &  \,0\, &  \,1\, & \cdots \\
    \vdots        & \vdots        & \vdots        & \vdots & \ddots & \ddots \\
    \end{array}\right\rceil ,
    \label{eq:bgeom}
\end{equation}
where $\delta_k=\check{\beta}_k-\check{\beta}_1^k$, $\check{\beta}_k=\beta_k/\beta_0$, with the nonuniversal $\beta_k$ ($k>1$) of
the original scheme.
The geometric scheme can be reached from any starting point: first introduce a scale change to align, say, $\Lambda_\text{lat}$ with
$\Lambda_{\MSbar}$; then the coefficient vector $\bm{r}_\text{g}=\mathbf{b}_\text{g}\cdot\bm{r}$ is independent of the ultraviolet
regulator and renormalization used to obtain~$\bm{r}$.

\section{One power correction}
\label{sec:MRS}

Let us recall how refs.~\cite{Komijani:2017vep,Brambilla:2017hcq} handle the pole mass.
The heavy-quark effective theory provides an expression for a heavy-light hadron mass~\cite{Falk:1992wt,Falk:1992fm,Mannel:1994kv}
along the lines of \cref{eq:R}:
\begin{equation}
    \mathcal{M} = \bar{m}\left(1 + \sum_{l=0} r_l \as^{l+1}(\bar{m})\right) + \bar\Lambda + \mathrm{O}(1/\bar{m}),
    \label{eq:M=m}
\end{equation}
where $\bar{m}=m_{\MSbar}(\mu)$ evaluated at $\mu=\bar{m}$, and $\bar{\Lambda}$, which is of order $\Lambda$, is the energy of
gluons and light quarks.
The series times $\bar{m}$ is known as the pole (or on-shell) mass.
The coefficients~$r_l$ are obtained from the quark self-energy by putting the quark on shell iteratively at each order in
perturbation theory.
The coefficients are infrared finite and gauge independent at every order of the iteration~\cite{Kronfeld:1998di}, but they grow
factorially with the order~$l$~\cite{Bigi:1993zi,Bigi:1994em,Beneke:1994sw,Beneke:1994rs}.
The series thus diverges, rendering its interpretation ambiguous.
A hadron mass cannot be ambiguous, so the ambiguity in the series must be canceled by $\bar\Lambda$ (and higher-power
terms)~\cite{Luke:1994xd}.

Komijani~\cite{Komijani:2017vep} exploited the fact that the leading factorial growth in the series, being related to~$\bar\Lambda$,
is independent of~$\bar{m}$.
Therefore, taking a derivative with respect to $\bar{m}$ generates a quantity without $\bar\Lambda$.
The derivative yields
\begin{equation}
    1 + \sum_{l=0} r_l \as^{l+1}(\bar{m}) + 2\beta\left(\as(\bar{m})\right)\sum_{l=0} (l+1) r_l \as^l(\bar{m}) \equiv
    1 + \sum_{k=0} f_k \as^{k+1}(\bar{m}),
    \label{eq:fr-mass}
\end{equation}
where the $f_k$ are obtained by expanding out $\beta(\as)$ on the left-hand side:
\begin{equation}
    f_k = r_k - 2\sum_{l=0}^{k-1} (l+1) \beta_{k-1-l} r_l.
    \label{eq:fk}
\end{equation}
\Cref{eq:fk} is eq.~(2.3) of ref.~\cite{Komijani:2017vep}.

Komijani recast \cref{eq:fr-mass,eq:fk} as a differential equation (eq.~(1.6) of ref.~\cite{Komijani:2017vep}),
\begin{equation}
    r(\alpha) + 2\beta(\alpha)r'(\alpha) = f(\alpha),
    \label{eq:1.6}
\end{equation}
where the prime denotes a derivative with respect to~$\alpha$.
The appendix of ref.~\cite{Komijani:2017vep} derives an asymptotic solution to \cref{eq:1.6} that pins down the normalization of the
large-order coefficients~$r_l$, $l\gg1$, i.e., the quantity denoted $K$ in \cref{sec:intro}.
Note that ref.~\cite{Komijani:2017vep} obtains a particular solution to \cref{eq:1.6}.
A general solution consists of any particular solution plus a solution to the corresponding homogeneous equation with $0$ instead of
$f(\alpha)$ on the right-hand side.
The solution of the homogeneous equation is a constant of order~$\Lambda$.
In this paper, \cref{eq:fk} is used instead of \cref{eq:1.6} as the starting point in search of a particular solution.

Before presenting the solution, let us generalize Komijani's idea to \cref{eq:R}: multiply $\mathcal{R}$ by $Q^p$ so the $\Lambda^p$
term no longer depends on~$Q$, differentiate once with respect to $Q$, and then divide by $pQ^{p-1}$:
\begin{equation}
    \mathcal{F}^{(p)}(Q) \equiv \hat{Q}^{(p)}\mathcal{R}(Q)
        \equiv \frac{1}{pQ^{p-1}}\frac{\dd\,Q^p\mathcal{R}}{\dd Q} = r_{-1} + F^{(p)}(Q) .
\end{equation}
In this case $F^{(p)}=\hat{Q}^{(p)}R$ also, and a nonzero $r_{-1}$ cancels out just like the $1$ in \cref{eq:fr-mass}.
Introducing a series for $F^{(p)}$ and collecting like powers of \as,
\begin{subequations} \label[subequations]{eq:fr}
\begin{equation}
    F^{(p)}(Q) = \sum_{k=0} f_k^{(p)}(\mu/Q) \as(\mu)^{k+1},
    \qquad
    f_k^{(p)} = r_k-\frac{2}{p}\sum_{l=0}^{k-1} (l+1) \beta_{k-1-l} r_l.
    \label{eq:fr-coeff}
\end{equation}
In matrix notation,
\begin{equation}
    \bm{f}^{(p)} = \mathbf{Q}^{(p)} \cdot \bm{r} , \qquad
    \mathbf{Q}^{(p)} = \mathbf{1} - \frac{2}{p}\mathbf{D} ,
    \label{eq:fr-mx}
\end{equation}
\end{subequations}
with $\mathbf{D}$ defined above.

\Cref{eq:fr} can be derived either by keeping $\as(\mu)$ independent of~$Q$ and taking the derivative of the coefficients or by
setting $\mu=Q$, as in \cref{eq:fr-mass}, so the coefficients are constant with $\as(Q)$ encoding the $Q$ dependence.
\Cref{eq:fr} generalize \cref{eq:fk,eq:1.6} to arbitrary~$p$; the differential equation \`a~la \cref{eq:1.6} corresponding to
\cref{eq:fr} has $2/p$ multiplying $\beta(\alpha)$.
The particular solution to the differential equation is simply obtained by solving \cref{eq:fr-mx}:
$\bm{r}={\mathbf{Q}^{(p)}}^{-1}\cdot\bm{f}^{(p)}$.

At this point, one might wonder what could be gained this way.
For some~$L$, the~$r_l$, $l<L$, are available in the literature.
Via \cref{eq:fr-coeff}, just as many $f_k^{(p)}$ are obtained from these $L$ terms and the first $L$ coefficients~$\beta_j$
(\cref{eq:beta}).
Solving \cref{eq:fr-mx} should just return the original information.
That is, of course, correct, but the solution, spelled out below, \emph{also} yields information about the $r_l$ for $l\ge L$.
Exploiting this additional information is the gist of this analysis.

The solution of \cref{eq:fr-mx} is easiest in the geometric scheme.
Let $b\equiv\beta_1/2\beta_0^2$, so that $2\beta_k=(2\beta_0)^{k+1} b^k$ (in the geometric scheme), and let $\tau\equiv2\beta_0/p$.
Then $\mathbf{Q}^{(p)}_\text{g}=\mathbf{b}_\text{g}\cdot\mathbf{Q}^{(p)}\cdot\mathbf{b}^{-1}_\text{g}$ has elements
\begin{subequations} \label[subequations]{eq:Q}
\begin{equation}
    \left\lceil Q^{(p)}_\text{g}\right\rceil_{kl} = \left\{ \begin{array}{ll}
         0,              & k<l, \\
         1,              & k=l, \\
        -(l+1)\tau^{k-l}(pb)^{k-l-1},\; & k>l , \\
    \end{array} \right.
    \label{eq:Qelem}
\end{equation}
which looks like
\begin{equation}
    \newcommand{\pb}{(\tau pb)}
    \mathbf{Q}^{(p)}_\text{g} = \left\lceil \begin{array}{ccccc@{\;\;}c@{\;\;}c@{\;\;}c}
             1      &      0      &      0      &      0      &      0     &    0    &    0   & \cdots \\
         -\tau      &      1      &      0      &      0      &      0     &    0    &    0   & \cdots \\
         -\tau^2pb  & -2\tau      &      1      &      0      &      0     &    0    &    0   & \cdots \\
         -\tau\pb^2 & -2\tau^2pb  & -3\tau      &      1      &      0     &    0    &    0   & \cdots \\
         -\tau\pb^3 & -2\tau\pb^2 & -3\tau^2pb  & -4\tau      &      1     &    0    &    0   & \cdots \\
         -\tau\pb^4 & -2\tau\pb^3 & -3\tau\pb^2 & -4\tau^2pb  & -5\tau     &    1    &    0   & \cdots \\
         -\tau\pb^5 & -2\tau\pb^4 & -3\tau\pb^3 & -4\tau\pb^2 & -5\tau^2pb &  -6\tau &    1   & \cdots \\
        \vdots      & \vdots      & \vdots      & \vdots      & \vdots     & \ddots  & \ddots & \ddots \\
    \end{array} \right\rceil .
    \label{eq:Bmx}
\end{equation}
\end{subequations}
$\mathbf{Q}^{(p)}_\text{g}$ exhibits geometric but not factorial growth.
\pagebreak
The inverse is easily obtained row-by-row:
\begin{subequations} \label[subequations]{eq:Q-1}
\begin{equation}
    \newcommand{\pb}[2]{\case{\Gamma(#1+pb)}{\Gamma(#2+pb)}} 
    \hspace*{-6pt}
    {\mathbf{Q}^{(p)}_\text{g}}^{-1} = \left\lceil \begin{array}{ccccc@{\quad}c@{\quad}c@{\quad}c}
           1      &     0      &    0   &    0   &  0 &  0 &  0 &  \cdots \\[0.5em]
        \tau      &     1      &    0   &    0   &  0 &  0 &  0 &  \cdots \\[0.5em]
        \tau^2\pb{3}{2} & 2\tau      &    1   &    0   &  0 &  0 &  0 & \cdots \\[0.5em]
        \tau^3\pb{4}{2} & 2\tau^2\pb{4}{3} & 3\tau &  1 & 0 &  0 &  0 & \cdots \\[0.5em]
        \tau^4\pb{5}{2} & 2\tau^3\pb{5}{3} & 3\tau^2\pb{5}{4} & 4\tau & 1 & 0 &  0 & \cdots \\[0.5em]
        \tau^5\pb{6}{2} & 2\tau^4\pb{6}{3} & 3\tau^3\pb{6}{4} & 4\tau^2\pb{6}{5} & 5\tau & 1 &  0 & \cdots \\[0.5em]
        \tau^6\pb{7}{2} & 2\tau^5\pb{7}{3} & 3\tau^4\pb{7}{4} & 4\tau^3\pb{7}{5} & 5\tau^2\pb{7}{6} & 6\tau & 1 & \cdots \\
        \vdots    & \vdots     & \vdots & \vdots & \vdots & \vdots & \ddots & \ddots \\
    \end{array} \right\rceil
    \label{eq:Q-1mx}
\end{equation}
or, expressed as in \cref{eq:Qelem},
\begin{equation}
    \left\lceil{Q^{(p)}_\text{g}}^{-1}\right\rceil_{lk} = \left\{ \begin{array}{ll}
        0,              & l<k, \\
        1,              & l=k, \\ \displaystyle
        (k + 1) \frac{\tau^{l} \Gamma(l + 1 + pb)}{\tau^{k} \Gamma(k + 2 + pb)}, & l>k. \\
    \end{array} \right.
    \label{eq:Q-1elem}
\end{equation}
\end{subequations}
From one row to the next, the entries increase both in a factorial way and by powers of~$\tau$.
As stated in \cref{sec:intro}, the growth starts at low orders.
From one column to the next, the entries \emph{decrease} factorially (and by powers of~$\tau$).
Both factorials grow rapidly only once $l\gg pb$, $k\gg pb$, so --- again as stated in \cref{sec:intro} --- the higher the
power~$p$, the longer the growth need not be apparent from explicit expressions for the coefficients.
Growth is also postponed for large~$b$, which happens if $\beta_0$ is small but $\beta_1$ is not.

Reexpressing \cref{eq:Q-1} as series coefficients,
\begin{equation}
    r_l = f_l^{(p)} + \left(\frac{2\beta_0}{p}\right)^l \Gamma(l+1 + pb)
        \sum_{k=0}^{l-1} \frac{k + 1}{\Gamma(k+2 + pb)} \left(\frac{p}{2\beta_0}\right)^k f_k^{(p)} ,
    \label{eq:rrenorm}
\end{equation}
which holds (in the geometric scheme) for all $l$.
\Cref{eq:rrenorm} is similar to eq.~(2.22) of ref.~\cite{Komijani:2017vep}, except for three details: eq.~(2.22) of
ref.~\cite{Komijani:2017vep} omits the first term $f_l^{(p)}$, has $\infty$ as the upper limit of the sum, and holds only
asymptotically (i.e., the relation is $\sim$ instead of $=$).
$\bm{f}^{(p)}$ grows more slowly than $\Gamma(l+1 + pb)$ or $\Gamma(k+2 + pb)$, so for $l\gg1$ it is accurate to neglect the first
term and to extend the sum to $\infty$.
The crucial difference is that \cref{eq:rrenorm} holds for all $l$, starting with the next few orders beyond the known~$r_l$.

Recall that $L$ terms are available.
Nowadays, $L=4$ for some problems (e.g., \cref{eq:M=m} and \cref{sec:E0}) and $L=3$ for others.
For $l<L$, \cref{eq:rrenorm} returns the~$r_l$ available at the outset.
For $l\ge L$, \cref{eq:rrenorm} suggests estimating $r_l$ (in the geometric scheme) by
\begin{subequations} \label[subequations]{eq:RlR0}
\begin{align}
    r_l \approx R_l^{(p)} &\equiv R_0^{(p)} \left(\frac{2\beta_0}{p}\right)^l \frac{\Gamma(l+1+pb)}{\Gamma(1+pb)} , \quad l\ge L ,
    \label{eq:rlRl} \\
    R_0^{(p)} &\equiv \sum_{k=0}^{L-1} (k + 1) \frac{\Gamma(1+pb)}{\Gamma(k+2+pb)} \left(\frac{p}{2\beta_0}\right)^k f_k^{(p)} .
    \label{eq:R0}
\end{align}
\end{subequations}
The expression for $R_0^{(1)}$ is the same as that for $N_{k_\text{max}}$ (with $k_\text{max}=L-1$) in eq.~(2.23) of
ref.~\cite{Komijani:2017vep}.
It is also resembles the formula (taken in the geometric scheme) for $P_{1/2}$ in eqs.~(17) of ref.~\cite{Hoang:2008yj}.
Applying \cref{eq:RlR0} to the series $R(Q)$ yields
\begin{subequations} \label[subequations]{eq:Rimproved}
\begin{equation}
    R(Q) \approx \sum_{l=0}^{L-1}  r_l   \ag^{l+1}(Q) + \sum_{l=L}^\infty R_l^{(p)} \ag^{l+1}(Q) .
    \label{eq:Rimproved-a}
\end{equation}
The first $L$ terms are as usual and the others are estimated via their fastest growing part.
For subsequent analysis, it is better to start the second sum at $l=0$,
\begin{equation}
    R(Q) \approx \sum_{l=0}^{L-1} \left(r_l-R_l^{(p)}\right) \ag^{l+1}(Q) + \sum_{l=0}^\infty R_l^{(p)} \ag^{l+1}(Q) ,
    \label{eq:Rimproved-b}
\end{equation}
\end{subequations}
which follows from subtracting and adding $\sum_{l=0}^{L-1}R_l^{(p)}\ag^{l+1}(Q)$.
For convenience below, let
\begin{equation}
    R_\text{RS}^{(p)}(Q) \equiv \sum_{l=0}^{L-1} \left(r_l-R_l^{(p)}\right) \ag(Q)^{l+1} , \qquad
    R_\text{B}^{(p)}(Q)  \equiv \sum_{l=0}^\infty R_l^{(p)} \ag(Q)^{l+1} .
    \label{eq:RRS}
\end{equation}
$R_\text{RS}^{(p)}$ is similar to the truncation to $L$ terms of the ``renormalon subtracted'' (RS) scheme for
$R$~\cite{Pineda:2001zq}.
Here, $R_\text{RS}^{(p)}$ arises not by intentional subtraction but from rearranging terms.
In the examples of the pole mass~\cite{Brambilla:2017hcq} and the static energy (\cref{sec:E0}), $r_l-R_l^{(p)}$ is smaller
than~$r_l$, especially for $l=3$,~$4$.

Because of the factorial growth of the $R_l^{(p)}$, the series $R_\text{B}^{(p)}$ does not converge.
It can be assigned meaning through Borel summation, however.
Using the integral representation of $\Gamma(l+1)$,
\begin{align}
    R_\text{B}^{(p)}(Q) &= R_0^{(p)} \sum_{l=0}^\infty \left[\frac{\Gamma(l+1 + pb)}{\Gamma(1 + pb)\Gamma(l+1)}
        \int_0^\infty \left(\frac{2\beta_0t}{p}\right)^l \e^{-t/\ag(Q)} \dd{t} \right] , \nonumber \\
        &\to R_0^{(p)} \int_0^\infty \frac{\e^{-t/\ag(Q)}}{(1-2\beta_0t/p)^{1+pb}} \dd{t} ,
    \label{eq:R-Borel}
\end{align}
where the second line is obtained by swapping the order of summation and integration.
Strictly speaking, the swap is not allowed because the integrand has a branch point at $t=p/2\beta_0$.
This singularity is known as a renormalon~\cite{tHooft:1977xjm}.
It is customary to place the cut on the real axis from the branch point to $+\infty$.
In ref.~\cite{Brambilla:2017hcq}, we split the integral into two parts, over the intervals $[0,p/2\beta_0
)$ 
before the cut and $[p/2\beta_0,\infty
)$ 
along the cut.
The first integral is unambiguous and given below.

For the interval $[p/2\beta_0,\infty
)$, 
the contour must be specified.
Taking it slightly above or below the cut, for example, yields
\begin{align}
    \delta R^{(p)} &\equiv R_0^{(p)} \int_{p/2\beta_0\pm\iunit\varepsilon}^{\infty\pm\iunit\varepsilon}
        \frac{\e^{-t/\ag(Q)}}{(1-2\beta_0t/p)^{1+pb}} \dd{t}
        = - R_0^{(p)} \e^{\pm\iunit pb\pi} \frac{p^{1+pb}}{2^{1+pb}\beta_0} \Gamma(-pb)
            \left[\frac{\e^{-1/[2\beta_0\ag(Q)]}}{[\beta_0\ag(Q)]^b} \right]^p ,
    \label{eq:R-Borel-ambiguous}
\end{align}
and the factor $\e^{\pm\iunit pb\pi}$ illustrates the ambiguity.
The quantity inside the bracket is identically $\Lambda_\text{g}/Q=\Lambda_{\MSbar}/Q$, so without loss
$\delta{}R^{(p)}\propto(\Lambda/Q)^p$ can be lumped into the solution of the homogeneous differential equation \`a~la \cref{eq:1.6}
or, equivalently, the power correction $C_p\Lambda^p/Q^p$ in \cref{eq:R}~\cite{Brambilla:2017hcq}.

Because the interchange of summation and integration in \cref{eq:R-Borel} is not allowed, $R_\text{B}^{(p)}$ can be \emph{assigned}
to be (taking $b<0$ at first and then applying analytic continuation)
\begin{subequations} \label[subequations]{eq:RB}
\begin{align}
    R_\text{B}^{(p)}(Q) &= R_0^{(p)} \int_0^{p/2\beta_0} \frac{\e^{-t/\ag(Q)}}{(1-2\beta_0t/p)^{1+pb}} \dd{t}
        = R_0^{(p)} \frac{p}{2\beta_0} \mathcal{J}(pb,1/2\beta_0\ag(Q)) ,
    \label{eq:Jfunction} \\
    \mathcal{J}(c,y) &= e^{-y} \Gamma(-c) \gamma^\star(-c,-y) ,
\end{align}
\end{subequations}
which is acceptable because the asymptotic (small \ag) expansion of $\mathcal{J}$ returns the original series in \cref{eq:RRS}.
Here $\gamma^\star(a,x)\equiv[1/\Gamma(a)]\int_0^1\dd{t}\,t^{a-1}\e^{-xt}$ is known as the limiting function of the incomplete gamma
function~\cite{AbramowitzStegun:1972}.
It is analytic in $a$ and~$x$ and has a convergent expansion
\begin{equation}
    \gamma^\star(a,-y) = \frac{1}{\Gamma(a)} \sum_{n=0}^\infty \frac{y^n}{n!(n+a)}, \;\; \forall y,
    \label{eq:gammaStar}
\end{equation}
which saturates quickly, also when $a=-pb<0$.

Combining the various ingredients leads to the prescription
\begin{equation}
    \mathcal{R}(Q) \equiv r_{-1} + R_\text{RS}^{(p)}(Q) + R_\text{B}^{(p)}(Q) + C_p\frac{\Lambda^p}{Q^p}
    \label{eq:RJ}
\end{equation}
for estimating $\mathcal{R}(Q)$.
Here, $R_\text{RS}^{(p)}(Q)$ is introduced in \cref{eq:RRS} and $R^{(p)}_\text{B}(Q)$ is defined by the right-hand side of
\cref{eq:Jfunction}.
\Cref{eq:RJ} is just eq.~(2.25) of ref.~\cite{Brambilla:2017hcq}, generalized to $p$ different from~$1$.

For the relation between the pole mass and \MSbar\ mass, ref.~\cite{Brambilla:2017hcq} referred to \cref{eq:RJ} as ``minimal
renormalon subtraction'' (MRS) in analogy with the RS mass of ref.~\cite{Pineda:2001zq}.
The derivation given here arguably does not subtract anything but instead adds new information to the usual truncated perturbation
series, rearranges a few terms, and then assigns meaning to an otherwise ill-defined series expression.
Even so, this paper continues to refer to the procedure as~MRS.
For example, it is often convenient to consider $R_\text{RS}^{(p)}(Q)+R_\text{B}^{(p)}(Q)\equiv R_\text{MRS}(Q)$ as a single
object.
The asymptotic (small \ag) expansion of $R_\text{MRS}(Q)$ is identical to the original series $R(Q)$.

Starting with \cref{eq:Rimproved}, the renormalization scale has been chosen to be $\mu=Q$.
If $\mu=sQ$ is chosen instead, the derivations do not change.
The coupling $\ag(Q)$ simply becomes $\ag(sQ)$ and the coefficients $r_l=r_l(1)$ and $f_k^{(p)}=f_k^{(p)}(1)$ become $r_l(s)$ and
$f_k^{(p)}(s)$.
How these effects play out in practice is discussed in \cref{sec:E0}.
In~$\delta R^{(p)}$, the bracket in \cref{eq:R-Borel-ambiguous} becomes $[\Lambda_\text{g}/sQ]^p$, so the overall change is to
replace $R_0^{(p)}(1)$ with $R_0^{(p)}(s)/s^p$.

\section{Cascade of power corrections}
\label{sec:FRS}

In general, problems like \cref{eq:R} have more than one power correction.
If there are two, with $p_2>p_1$, $\mathcal{F}^{{p_1}}$ still contains $(p_1-p_2)C_{p_2}\Lambda^{p_2}/p_1Q^{p_2}$, which can be
removed with $\hat{Q}^{(p_2)}$:
\begin{equation}
    \bm{f}^{\{p_1,p_2\}} \equiv \mathbf{Q}^{(p_2)}       \cdot \bm{f}^{(p_1)} \quad\Rightarrow\quad
    \bm{f}^{(p_1)}        =    {\mathbf{Q}^{(p_2)}}^{-1} \cdot \bm{f}^{\{p_1,p_2\}}.
    \label{eq:f-p1p2}
\end{equation}
These coefficients could then be used in \cref{eq:rrenorm}.
A similar idea was mentioned in v1 and v2 on \href{https://arXiv.org/abs/1701.00347v2}{arXiv.org} of ref.~\cite{Komijani:2017vep}.
With the early onset of the ``large-$l$'' behavior not yet clear when ref.~\cite{Komijani:2017vep} was written, the utility of
\cref{eq:f-p1p2} was also not clear.
For whatever reason, the discussion was removed from the final publication.

More concretely and in general, if the set of powers is $\{p_1,p_2,\ldots,p_n\}$, the operator (with $\hat{Q}^{(p_1)}$ rightmost)
\begin{equation}
    \hat{Q}^{\{p_i\}} = \prod_{j=0}^{n-1} \hat{Q}^{(p_{n-j})}
\end{equation}
fully removes the power corrections associated with these powers.
In matrix notation, the $F$-series coefficients
\begin{equation}
    \bm{f}^{\{p_i\}} = \mathbf{Q}^{\{p_i\}} \cdot \bm{r} = \prod_{j=0}^{n-1} \mathbf{Q}^{(p_{n-j})} \cdot \bm{r}
\end{equation}
are obtained with $\mathbf{Q}^{\{p_i\}}$, which is the obvious matrix representation of $\hat{Q}^{\{p_i\}}$.
This equation can be solved for
\begin{equation}
    \bm{r} = {\mathbf{Q}^{\{p_i\}}}^{-1} \cdot \bm{f}^{\{p_i\}} =
        \prod_{j=1}^{n} {\mathbf{Q}^{(p_j)}}^{-1} \cdot \bm{f}^{\{p_i\}},
\end{equation}
and, as above, the series $R(Q)$ is approximated by using the $L$ known terms of $\bm{r}$ while using the rest of them from this
solution.

Because the $\mathbf{Q}^{(p_i)}$ commute, their inverses do, so a partial-fraction decomposition turns the product into a sum,
\begin{equation}
    \prod_{j=1}^n {\mathbf{Q}^{(p_j)}}^{-1} = \sum_{j=1}^n h_j^{\{p_i\}} {\mathbf{Q}^{(p_j)}}^{-1} , \qquad
    h_j^{\{p_i\}} = \prod_{k=1,k\neq j}^n \frac{p_k}{p_k-p_j}.
    \label{eq:product=sum}
\end{equation}
Note that $\sum_{j=1}^nh_j^{\{p_i\}}=1$, $\sum_{j=1}^np_jh_j^{\{p_i\}}=0$; \cref{tab:partition} shows the $h_j^{\{p_i\}}$ for
various sets~$\{p_i\}$.
\begin{table}
\centering
\begin{tabular}{|c|cccccc|}
\hline
$\{p_i\}\;\setminus\;\;j$ &   $1$  &  $2$ &  $3$ &  $4$ &  $6$ &  $8$ \\
\hline
\{1,2\}     &   $2$  & $-1$ &   --   &   --  &   --   &  --  \\
\{1,3\}     &  $3/2$ &  --  & $-1/2$ &   --  &   --   &  --  \\
\{1,2,3\}   &   $3$  & $-3$ &  $1$   &   --  &   --   &  --  \\
\{1,2,4\}   &  $8/3$ & $-2$ &   --   & $1/3$ &   --   &  --  \\
\{1,2,3,4\} &   $4$  & $-6$ &  $4$   &  $-1$ &   --   &  --  \\
\{2,4\}     &   --   & $2$  &   --   &  $-1$ &   --   &  --  \\
\{2,4,6,8\} &   --   &  $4$ &   --   &  $-6$ &  $4$   & $-1$ \\
\{4,6\}     &   --   &  --  &   --   &  $3$  &  $-2$  &  --  \\
\{4,6,8\}   &   --   &  --  &   --   &  $6$  &  $-8$  &  $3$ \\
\{1,2,4,6\} & $16/5$ &  --  &  $-3$  &  $1$  & $-1/5$ &  --  \\
\{1,3,4,6,8\}&$96/35$&  --  & $-32/5$&  $6$  & $-8/5$ & $9/35$ \\
\hline
\end{tabular}
\caption{Partition coefficients $h_j^{\{p_i\}}$ for various sets of powers $p_i$.}
\label{tab:partition}
\end{table}
The solution is thus,
\begin{equation}
    \bm{r} = \sum_{j=1}^n h_j^{\{p_i\}} {\mathbf{Q}^{(p_j)}}^{-1} \cdot \bm{f}^{\{p_i\}} .
    \label{eq:r-multi}
\end{equation}
which generalizes \cref{eq:rrenorm}.
The prescription is again to take the first $L$ $r_l$ as computed in the literature and approximate the rest with the leading
factorials in \cref{eq:r-multi}.
That means
\begin{subequations} \label[subequations]{eq:FRS}
\begin{align}
    \mathcal{R}(Q) &\equiv r_{-1} + R_\text{RS}^{(p)}(Q) + R_\text{B}^{(p)}(Q) + \sum_{i=1}^n C_{p_i}\frac{\Lambda^{p_i}}{Q^{p_i}} ,
    \label{eq:RJJ} \\
    R_\text{RS}(Q) &\equiv \sum_{l=0}^{L-1} \left(r_l-R_l^{\{p_i\}}\right) \ag^{l+1}(Q)  , \\
    R_\text{B}(Q)  &\equiv \sum_{j=1}^n h_j^{\{p_i\}} R_0^{\{p_i\}(p_j)} \frac{p_j}{2\beta_0}
        \mathcal{J}\left(p_jb,p_j/2\beta_0\ag(Q)\right) ,
\end{align}
where
\begin{align}
    R_l^{\{p_i\}} &=      \sum_{j=1}^n h_j^{\{p_i\}} R_0^{\{p_i\}(p_j)}
        \left(\frac{2\beta_0}{p_j}\right)^l \frac{\Gamma(l+1 + p_jb)}{\Gamma(1 + p_jb)} , \\
    R_0^{\{p_i\}(p_j)} &= \sum_{k=0}^{L-1} (k + 1) \frac{\Gamma(1 + p_jb)}{\Gamma(k+2 + p_jb)}
        \left(\frac{p_j}{2\beta_0}\right)^k f_k^{\{p_i\}} .
\end{align}
\end{subequations}
The same $f_k^{\{p_i\}}$ appear in all $R_0^{\{p_i\}(p_j)}$, hence the somewhat fussy notation.
For lack of a better name, MRS now stands for ``multiple renormalon subtraction'' even though, again, the procedure as developed
here adds information.
A possible notation to distinguish how many power corrections have been removed from a given series is
``MRS$\{p_1,p_2,\ldots,p_n\}$''.

\section{Other renormalization schemes}
\label{sec:scheme}

While the geometric scheme simplifies the solution of the matrix equations, it is useful to generalize MRS to arbitrary schemes.
Given the algebra of \cref{sec:MRS}, the simplest way to solve to \cref{eq:fr-mx} is to combine \cref{eq:bgeom,eq:Q-1mx}, yielding
\begin{equation}
    {\mathbf{Q}^{(p)}}^{-1} = \mathbf{b}^{-1}_\text{g}\cdot{\mathbf{Q}^{(p)}_\text{g}}^{-1}\cdot\mathbf{b}_\text{g} =
        {\mathbf{Q}^{(p)}_\text{g}}^{-1} + \mathbf{\Delta}^{(p)}.
    \label{eq:Qinv}
\end{equation}
The lower-triangular matrix $\mathbf{\Delta}^{(p)}$ contains the $\delta_k$ introduced immediately after \cref{eq:bgeom}, which
parametrize the deviation from the geometric scheme of the arbitrary-scheme $\beta$-function coefficients.
The same result is obtained, of course, by solving \cref{eq:fr-mx} directly and eliminating the $\beta_k$ in favor of
the~$\delta_k$.

Another way to express \cref{eq:Qinv} is
\begin{equation}
    {\mathbf{Q}^{(p)}}^{-1} = {\mathbf{Q}^{(p)}_\text{g}}^{-1}\left(\mathbf{1} + \mathbf{K}^{(p)}\right),
    \label{eq:Qinv-K}
\end{equation}
where $\mathbf{K}^{(p)}$ looks like
\begin{equation}
    \newcommand{\cb}{\tau}
    \newcommand{\cu}{2\cb^2\delta_2 + \cb\delta_3}
    \newcommand{\ct}{3\tau^3(2+pb)\delta_2 + 2\tau^2\delta_3 + \tau\delta_4}
    \newcommand{\cw}{6\cb^2\delta_2 + 2\cb\delta_3}
    \mathbf{K}^{(p)} = \left\lceil\begin{matrix}
              0     &       0      &       0      & \;0\;  & \;0\;  & \;0\;  \\
              0     &       0      &       0      &    0   &    0   &    0   \\
              0     &       0      &       0      &    0   &    0   &    0   \\
        \cb\delta_2 &       0      &       0      &    0   &    0   &    0   \\
             \cu    & 2\cb\delta_2 &       0      &    0   &    0   &    0   \\
             \ct    &      \cw     & 3\cb\delta_2 &    0   &    0   &    0   \\
           \vdots   &    \vdots    &    \vdots    & \ddots & \vdots & \vdots
    \end{matrix}\right\rceil .
    \label{eq:K-matrix}
\end{equation}
The matrix $\mathbf{K}^{(p)}$ can be decomposed into matrix coefficients of $\delta_i$, $\delta_i\delta_j$, etc.
The matrix multiplying single powers of $\delta_i$ possess an easily seen pattern:
\begin{equation}
    \left. \frac{\partial K_{lk}}{\partial\delta_i}\right|_{\forall j,\delta_j=0} = \left\{\begin{array}{ll}
            0 ,      & l < k+i+1 \\
            (k + 1)         \tau , & l = k+i+1 \\  \displaystyle
            (k + 1) (l - i) \frac{\tau^{l-i} \Gamma(l - i + p b])}{\tau^k \Gamma(k + 2 + p b)} , & l > k+i+1 .
        \end{array} \right.
\end{equation}
For example, the term $3\tau^3(2+pb)\delta_2$ in $K^{(p)}_{50}$ is the first nontrivial term.
Starting on the $l=6$ row (not shown in \cref{eq:K-matrix}), $\mathbf{K}^{(p)}$ contains pieces proportional to $\delta_i^2$;
similarly, starting on the $l=7$ row, $\mathbf{K}^{(p)}$ contains pieces proportional to $\delta_i\delta_j$.
In neither case is any pattern to the matrix coefficient apparent.

The original correction $\mathbf{\Delta}^{(p)}$ looks similar to the right-hand side of \cref{eq:K-matrix}, but its structure, which
is most easily constructed from ${\mathbf{Q}^{(p)}_\text{g}}^{-1}\cdot\mathbf{K}^{(p)}$, is less illuminating than
$\mathbf{K}^{(p)}$'s.
The terms in $r_l$, $l\ge L$, stemming from $\mathbf{\Delta}^{(p)}$ are smaller than those from ${\mathbf{Q}^{(p)}_\text{g}}^{-1}$.
In previous work on the large-$l$ behavior of the $r_l$~\cite{Beneke:1994rs,Beneke:1998ui,Ayala:2014yxa,Komijani:2017vep}, the
$\delta_j$ appear in a way that does not look like the medium-$l$ pattern accessible by the matrix derivation.

In practice, however, the details of $\mathbf{K}^{(p)}$ may not matter.
Only the first few $\delta_j$ are known.
In the geometric scheme they enter the coefficients~$\bm{r}_\text{g}$ and~$\bm{f}_\text{g}$.
Thus, they may as well be absorbed into~$\bm{r}_\text{s}$ and~$\bm{f}_\text{s}$ by introducing
\begin{equation}
    \bm{f}_{\text{Ks}}^{(p)} \equiv \left(\mathbf{1}+\mathbf{K}^{(p)}\right) \cdot \bm{f}_\text{s}^{(p)}, \qquad
    \boldsymbol{\mathfrak{A}}_\text{s}\cdot\bm{r}_\text{s} =
        \boldsymbol{\mathfrak{A}}_\text{s}\cdot{\mathbf{Q}^{(p)}_\text{g}}^{-1}\cdot\bm{f}_{\text{Ks}}^{(p)}.
    \label{eq:fKs}
\end{equation}
Then Borel summation can be applied by combining the growing part of ${\mathbf{Q}^{(p)}_\text{g}}^{-1}$ with
$\boldsymbol{\mathfrak{A}}_\text{s}$ and combining the diminishing part of ${\mathbf{Q}^{(p)}_\text{g}}^{-1}$ with
$\bm{f}_{\text{Ks}}^{(p)}$ to form the normalization factor.
Indeed, if $L$ orders are available, and the scheme is chosen so that $\delta_j=0$ for all $j\le L-2$, then the upper-left
$L\times L$ block of $\mathbf{K}^{(p)}$ vanishes, and the knowable part of~$\bm{f}_{\text{Ks}}^{(p)}$ coincides
with~$\bm{f}_\text{s}^{(p)}$.

A reason to consider schemes other than the geometric coupling is that $\ag(\mu)$ runs into a branch point of the Lambert-$W$
function~\cite{Corless:1993lwf} at $\mu=(\e/2b)^b\Lambda$.
(For $N_c=n_f=3$, $(\e/2b)^b\approx1.629$.) %
\begin{figure}
    \centering
    \includegraphics[width=0.75\textwidth]{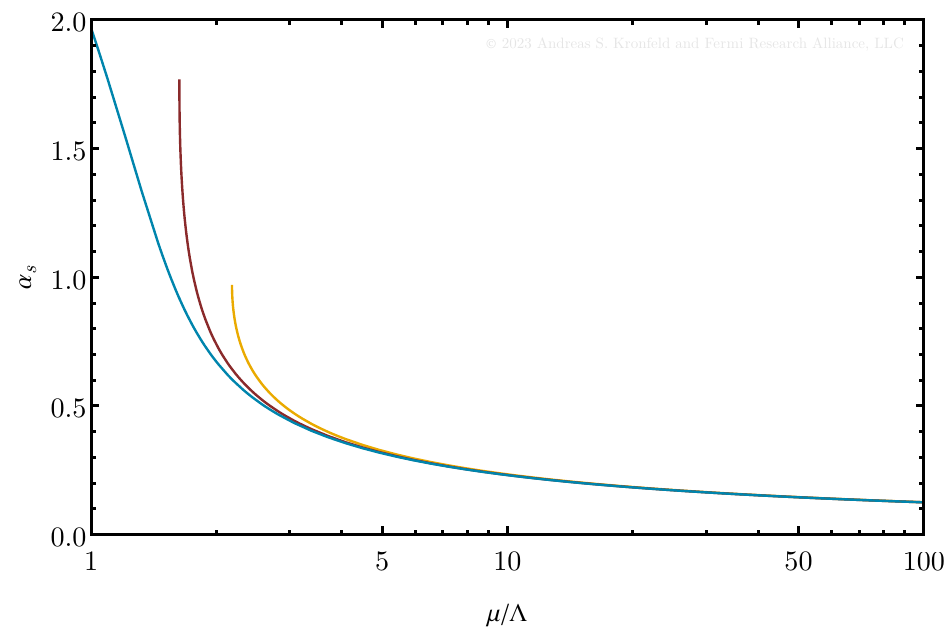}
    \caption[fig:alphaGeom]{Gauge coupling $\an{s}$ vs.\ $\mu/\Lambda$ in various schemes for $N_c=n_f=3$:
        $\an{\MSbar}(\mu)$~(gold), $\ag(\mu)$~(red), and $\an{2}(\mu)$~(blue).
        $\an{\MSbar}$ (with $1/\beta(\alpha)$ expanded to fixed order) and $\ag$ run into branch points at
        $\mu\approx2.1797\Lambda$ and $(\e/2b)^b\Lambda\approx1.629\Lambda$, respectively.
        For smaller $\mu$, they are undefined.
        At these points, $\an{\MSbar}(2.1797\Lambda)=0.97601$, $\ag(1.629\Lambda)=1.76715$.
        $\an{2}(\mu)$ does not behave this way and like \ag\ has $\delta_2=0$.}
    \label{fig:alphaGeom}
\end{figure}
\Cref{fig:alphaGeom} shows the running of \ag\ and \an{2} ($\an{n}$ for $n=2$), in SU(3) gauge theory with three massless flavors.
The pole in the geometric $\beta$~function, which is the source of the problem, can be removed while retaining a closed-form
relation between $\ln(\mu/\Lambda)$ and a family of schemes $\an{n}$:
\begin{subequations} \label[subequations]{eq:alpha-new}
\begin{align}
    \beta(\an{n}) &= -\frac{\beta_0\an{n}^2}{1-(\beta_1/\beta_0)\an{n}+n(\beta_1\an{n}/\beta_0)^{n+1}} ,
    \label{eq:beta-n} \\
    \ln(\mu/\Lambda) &= \frac{1}{2\beta_0\an{n}} + b \ln(\beta_0\an{n}) - b \left(\frac{\beta_1\an{n}}{\beta_0}\right)^n .
    \label{eq:alpha-n}
\end{align}
\end{subequations}
%
%
In \cref{sec:E0}, $\an{2}$ is used to study how MRS works in practice.
Like $\ag$, $\an{2}$ has $\delta_2=0$, so that $\mathbf{K}^{(p)}$ can be neglected (for $L\le4$).
\abar\ can be formulated by integrating the \MSbar\ $\beta$~function with either $1/\beta(\as)$ or $\beta(\as)$ expanded to fixed
order.
Both have an undesirable fixed point \`a~la \ag.
Truncating with $\beta_3$, the former choice --- also used in \cref{sec:E0} --- is valid only for $\mu\ge2.1797\Lambda$, at which
point $\as=0.97601$ (cf., \cref{fig:alphaGeom}).
The latter (again truncating with $\beta_3$) is valid only for $\mu\ge0.87645\Lambda$, asymptotically as $\as\to\infty$ and in
practice for $\as\gtrsim50$.


\section{Anomalous dimensions}
\label{sec:anom}

The $Q$ dependence is not always as simple as the power law in \cref{eq:R}, because $C_p$ can depend on $Q$ via $\as(Q)$.
In the operator-product expansion, for example, power corrections take the form
\begin{equation}
    C_p(\mu/Q, \as(\mu)) \frac{\langle \mathcal{O}(\mu) \rangle}{Q^p} =
        \widehat{C}_p(\as(Q)) \frac{\langle \mathcal{O}_\text{RGI} \rangle}{Q^p}.
\end{equation}
On the right-hand side, the renormalization group has been used to factor the $\mu$ dependence, such that
$\langle\mathcal{O}_\text{RGI}\rangle\propto\Lambda^p$.
The renormalization-group-invariant (RGI) Wilson coefficient can be written
\begin{equation}
    \widehat{C}_p(\as) = \left(2\beta_0\as\right)^\psi \sum_{l=-1} c_l \as^{l+1},
    \label{eq:wideC}
\end{equation}
where $\psi=\gamma_0/2\beta_0$ and $\gamma_0$ is the one-loop anomalous dimension of~$\mathcal{O}$.
Some of the leading coefficients may (for some reason) vanish, and the series is known in practice only to some order.
Strategies for truncating the series in \cref{eq:wideC} lie beyond the scope of this paper.

Let us assume $c_{-1}\neq0$.
It is convenient to extend the matrix notation to %
$\boldsymbol{\mathfrak{A}}_\text{g}=\left\lceil 1 \quad \ag \quad \ag^2 \quad \ag^3 \quad \ag^4 \quad \cdots \; \right\rceil$,
$\bm{r}_\text{g}=\left\lceil r_{-1} \quad r_0 \quad r_1 \quad r_2 \quad r_3 \quad \cdots \; \right\rceil^\text{T}$, and so on.
The $r_{-1}$ entry is useful for bookkeeping; it cannot influence the final result, so below it can be set to~0, which is equivalent
to changing to physical quantity to $\mathcal{R}(Q)-r_{-1}$.

To isolate $\Lambda^p$ so that it can be differentiated away, it is necessary to multiply by $Q^p/\widehat{C}_p$.
Division by the Wilson coefficient changes the series $\boldsymbol{\mathfrak{A}}_\text{g}\cdot\bm{r}_\text{g}$ to
$(2\beta_0\as)^{-\psi}\boldsymbol{\mathfrak{A}}_\text{g}\cdot\mathbf{C}^{-1}\cdot\bm{r}_\text{g}$, where
\begin{equation}
    \mathbf{C} = \left\lceil \begin{array}{ccccc}
        c_{-1} &   0    &   0    &   0    & \cdots \\
          c_0  & c_{-1} &   0    &   0    & \cdots \\
          c_1  &   c_0  & c_{-1} &   0    & \cdots \\
          c_2  &   c_1  &   c_0  & c_{-1} & \cdots \\
        \vdots & \vdots & \ddots & \ddots & \ddots
    \end{array}
    \right\rceil.
\end{equation}
The operation $\hat{Q}^{(p)}$ is applied to
$(2\beta_0\as)^{-\psi}\boldsymbol{\mathfrak{A}}_\text{g}\cdot\mathbf{C}^{-1}\bm{r}_\text{g}$, followed by multiplication by
$\widehat{C}_p$.
These steps yield $\boldsymbol{\mathfrak{A}}_\text{s}\cdot\bm{f}^{(p,\psi,\widehat{C})}_\text{g}$ with
\begin{equation}
    \bm{f}^{(p,\psi,\widehat{C})}_\text{g} =
        \mathbf{C} \cdot \mathbf{Q}^{(p,\psi)}_\text{g} \cdot \mathbf{C}^{-1} \cdot \bm{r}_\text{g},
    \label{eq:hatCpQ}
\end{equation}
and $\mathbf{Q}^{(p,\psi)}_\text{g}$ has the same entries as in \cref{eq:Q} but with $l\to l-\psi$ and $k\to k-\psi$.
The inverse ${\mathbf{Q}^{(p,\psi)}_\text{g}}^{-1}$ is given by \cref{eq:Q-1} with the same substitutions.

\Cref{eq:hatCpQ} can be solved for
\begin{equation}
    \bm{r}_\text{g} =
        \mathbf{C} \cdot {\mathbf{Q}^{(p,\psi)}_\text{g}}^{-1} \cdot \mathbf{C}^{-1} \cdot \bm{f}_\text{g}^{(p,\psi,\widehat{C})} ,
\end{equation}
which has the same structure as the scheme change \cref{eq:Qinv}.
Thus,
\begin{equation}
    \mathbf{C}\cdot{\mathbf{Q}^{(p,\psi)}_\text{g}}^{-1}\cdot\mathbf{C}^{-1}=
        {\mathbf{Q}^{(p,\psi)}_\text{g}}^{-1}\cdot\left(\mathbf{1} + \mathbf{K}^{(p,\psi,\widehat{C})} \right),
    \label{eq:CQC}
\end{equation}
and $\mathbf{K}^{(p,\psi,\widehat{C})}$ can be absorbed into the coefficients $\bm{f}_\text{g}^{(p,\psi,\widehat{C})}$, as in
\cref{eq:fKs} when estimating $r_l$, $l\ge L$.
In the basic formulas for the improved series, \cref{eq:RlR0}, it makes sense (along with $l\to l-\psi$ and $k\to k-\psi$) to change
the conventional factor $\Gamma(1+pb)$ to $\Gamma(1+pb-\psi)$ and to omit $\psi$ in the powers of $2\beta_0/p$.
The change to the normalization factor, $R_0^{(p,\psi)}$, is straightforward.
In the Borel summation leading up to \cref{eq:RB}, $\psi$ always appears as $pb-\psi$; the sum over $l$ and splitting of the
integration follow exactly as in \cref{sec:MRS}.

If more than one power correction has an anomalous dimension, they still can be removed successively.
Now every step affects all subsequent steps.
The case of removing two power corrections reveals how complications ensue.
Let the two power terms be $\widehat{C}_i\Lambda^{p_i}/Q^{p_i}$, $i=1,2$.
The first step converts the second Wilson coefficient
\begin{subequations} \label[subequations]{eq:C21}
\begin{align}
    \widehat{C}_2 &= (2\beta_0\as)^{\psi_2}\boldsymbol{\mathfrak{A}}_\text{s}\cdot \bm{c}_2 \mapsto
        \frac{p_1-p_2}{p_1} (2\beta_0\as)^{\psi_2} \boldsymbol{\mathfrak{A}}_\text{s}\cdot \bm{c}_{2/1} , \\
    \bm{c}_{2/1} &= \mathbf{C}_1 \cdot \mathbf{Q}^{(p_1-p_2,\psi_1-\psi_2)} \cdot \mathbf{C}_1^{-1} \cdot \bm{c}_2 .
\end{align}
\end{subequations}
The second step then leads to
\begin{equation}
    \bm{f}^{\{(p_1,\psi_1,\widehat{C})_1,(p_2,\psi_2,\widehat{C})_2\}}_\text{g} =
        \mathbf{C}_{2/1} \cdot \mathbf{Q}^{(p_2,\psi_2)}_\text{g} \cdot \mathbf{C}_{2/1}^{-1} \cdot
        \mathbf{C}_1     \cdot \mathbf{Q}^{(p_1,\psi_1)}_\text{g} \cdot \mathbf{C}_1^{-1} \cdot
        \bm{r}_\text{g},
    \label{eq:ugh}
\end{equation}
Note that the same outcome is obtained if the $\Lambda^{p_2}$ term is removed first, i.e.,
\begin{equation}
    \mathbf{C}_{2/1} \cdot \mathbf{Q}^{(p_2,\psi_2)}_\text{g} \cdot \mathbf{C}_{2/1}^{-1} \cdot
        \mathbf{C}_1 \cdot \mathbf{Q}^{(p_1,\psi_1)}_\text{g} \cdot \mathbf{C}_1^{-1} =
    \mathbf{C}_{1/2} \cdot \mathbf{Q}^{(p_1,\psi_1)}_\text{g} \cdot \mathbf{C}_{1/2}^{-1} \cdot
        \mathbf{C}_2 \cdot \mathbf{Q}^{(p_2,\psi_2)}_\text{g} \cdot \mathbf{C}_2^{-1} ,
    \label{eq:QQQQ}
\end{equation}
and similarly for their inverses.
A~decomposition of the right-hand side of \cref{eq:ugh} along the lines of \cref{eq:product=sum} seems possible by isolating
$\mathbf{Q}^{(p_1,\psi_1)}_\text{g}$ and $\mathbf{Q}^{(p_2,\psi_2)}_\text{g}$ and pragmatically absorbing the rest into the
coefficients (as in \cref{eq:CQC}), but an elegant arrangement has (so far) eluded me.

Suppose the $c_l$ vanish for $l<n$.
The first nonzero term, $c_n$, should not be connected to the $r_l$, $l\le n$.
A possible route forward is to subtract $\sum_{l=-1}^nr_l\as^{l+1}$ from $\mathcal{R}$, and the difference is still a valid
observable.
The factorially growing contributions can then be treated as before.
If $p_2=p_1$ but $\psi_2\neq\psi_1$, the vector $\bm{c}_{2/1}$ in \cref{eq:C21} must be redefined as
$-2\mathbf{C}_1\cdot\mathbf{D}^{(\psi_1-\psi_2)}\cdot\mathbf{C}_1^{-1}\cdot\bm{c}_2$ with $c_{2/1,-1}=0$, so the second step will
have to be tweaked in a similar way.

\section{The static energy}
\label{sec:E0}

To see MRS in action, the procedure is applied in this section to the gluonic energy stored between a static quark and a static
antiquark, $E_0(r)$, called the ``static energy'' for short.
It is computed in lattice gauge theory from the exponential fall-off at large~$t$ of a $t\times r$ Wilson
loop~\cite{Wilson:1974sk,Brambilla:2022het}.
The lattice quantity is the sum of a physical quantity plus twice the linearly divergent self-energy of a static quark.
Dimensional regularization has no linear divergence, but on general grounds a constant of order~$\Lambda$ is possible.
Setting $\mathcal{R}(1/r)=-rE_0(r)/C_F$ yields a quantity of the form given in \cref{eq:R} with $r_{-1}=0$ and $p=1$.

The static energy is a good candidate to test MRS because four orders in perturbation theory are known, thus enabling a thorough
test.
Beyond the tree-level result of order~\as, \MSbar-scheme results are available at order
$\as^2$~\cite{Fischler:1977yf,Billoire:1979ih}, $\as^3$~\cite{Peter:1997me,Schroder:1998vy,Kniehl:2001ju}, and
$\as^4$~\cite{Smirnov:2008pn,Anzai:2009tm,Smirnov:2009fh,Lee:2016cgz}.
The one-loop~\cite{Gross:1973id,Politzer:1973fx}, two-loop~\cite{Jones:1974mm,Caswell:1974gg},
three-loop~\cite{Tarasov:1980au,Larin:1993tp}, and four-loop~\cite{vanRitbergen:1997va,Czakon:2004bu,Zoller:2016sgq} coefficients of
the \MSbar\ $\beta$~function are also needed.
The five-loop coefficient~$\beta_4$~\cite{Baikov:2016tgj,Herzog:2017ohr,Luthe:2017ttc} is not needed here.

References~\cite{Fischler:1977yf,Billoire:1979ih,Peter:1997me,Schroder:1998vy,Kniehl:2001ju,Smirnov:2008pn,%
Anzai:2009tm,Smirnov:2009fh,Lee:2016cgz} compute the static potential, $V(q)$, in momentum space, finding it to be infrared
divergent starting at order~$\as^4$~\cite{Appelquist:1977es}.
This behavior reflects the emergence of an ``ultrasoft'' scale $\as r^{-1}$ in addition to the hard scale $r^{-1}$.
Ultrasoft contributions can be described in a multipole expansion and thereby demonstrated to render the static energy infrared
finite~\cite{Brambilla:1999qa,Brambilla:1999xf,Kniehl:1999ud}.
If $\as r^{-1}\gg\Lambda$, the ultrasoft part can be calculated perturbatively~\cite{Brambilla:1999qa,Kniehl:1999ud}, and the total
static energy is explicitly seen to be infrared finite~\cite{Brambilla:1999qa,Kniehl:1999ud,Anzai:2009tm}.
A~remnant of the cancellation remains in logarithms of the ratio of the two scales, $\ln[(\as r^{-1})/r^{-1}]=\ln\as$.

Following the exposition of Garcia i~Tormo~\cite{Tormo:2013tha}, a momentum-space quantity, here denoted~$\tilde{\mathcal{R}}(q)$,
poses a second problem \`a~la \cref{eq:R}, again with $r_{-1}=0$ but now with $p>1$.
To distinguish the series coefficients associated with $\tilde{\mathcal{R}}(q)$ and $\mathcal{R}(1/r)$ from each other and the
distance $r$, the notation used here is
\begin{equation}
    \tilde{R}(q) = \sum_{l=0} a_l(\mu/q) \as(\mu)^{l+1} , \quad
    \quad R(1/r) = \sum_{l=0} v_l(\mu r) \as(\mu)^{l+1} .
    \label{eq:static-series}
\end{equation}
The coefficients $a_l(1)$ are available in the literature~\cite{Fischler:1977yf,Billoire:1979ih,Peter:1997me,Schroder:1998vy,%
Kniehl:2001ju,Smirnov:2008pn,Anzai:2009tm,Smirnov:2009fh,Lee:2016cgz} and can be found in a consistent notation in the accompanying
Mathematica~\cite{Mathematica:13.2.1} notebook.
Each $v_l(\mu r)/r$ is $4\pi$ times the Fourier transform of $a_l(\mu/q)/q^2$.
Indeed, the $p=1$ factorial growth of the $v_l$ arises from the Fourier transform of the logarithms (cf., \cref{eq:r}) in
$a_l(\mu/q)$.
The series $F^{(1)}(1/r)$, derived as in \cref{sec:MRS} from $R(1/r)$, is related to the ``static force'',
$\mathfrak{F}(r)=-\dd{E_0}/\dd r$, by $\mathcal{F}(r)=F^{(1)}(1/r)=-r^2\mathfrak{F}(r)/C_F$.
Note that $\mathfrak{F}(r)$ --- and, hence $\mathcal{F}(r)$ and $F^{(1)}(1/r)$ --- is expected to be free of renormalon
ambiguities~\cite{Brambilla:1999xf,Ayala:2020odx}, because the change in static energy from one distance to another is physical.
The series $f_l$ should eventually exhibit factorial growth owing to instantons, i.e., with
$p\ge4\pi\beta_0=\case{11}{3}C_A-\case{4}{3}\sum_fT_f$.

The remainder of this section gives numerical and graphical results for SU(3) gauge theory with three massless flavors.
For brevity, the superscript ``(1)'' on $F^{(1)}$, $R_0^{(1)}$, etc., is omitted.
To obtain numerical results and prepare plots, \as\ in the ultrasoft logarithm, $\ln\as$, must be specified.
This $\as$ can be taken to run, namely taken to be the same as the expansion parameter~$\as(\mu)$.
Alternatively, $\as$ can be held fixed.
Below, $\as(s/r)$ (or $\as(sq)$), for various fixed $s$ is used as an expansion parameter, and the ultrasoft $\as$ is chosen either
to be the same or, for comparison, a fixed value $\as=\third$.
This value arises at scales where perturbation theory starts to break down, making it a reasonable alternative.
Resummation of logarithms $\as^{3+n}\ln^n\as$~\cite{Pineda:2000gza} and $\as^{4+n}\ln^n\as$~\cite{Brambilla:2006wp,Brambilla:2009bi}
is not considered here.

\Cref{tab:coeff-vf} shows the first four $a_l$ and $f_l$ in three different renormalization schemes, \MSbar, geometric, and
\cref{eq:alpha-new} with $n=2$.
\begin{table}
    \centering
    \sisetup{table-format = 3.6, table-alignment-mode = format}
    \begin{tabular}{|c|SS|SS|SS|}
    \hline
    & \multicolumn{2}{c|}{\MSbar} & \multicolumn{2}{c|}{geometric} & \multicolumn{2}{c|}{\cref{eq:alpha-new}, $n=2$} \\
    $l$ & {$a_l(1)$} & {$f_l(1)$} & {$a_l(1)$} & {$f_l(1)$} & {$a_l(1)$} & {$f_l(1)$} \\
    \hline
    $0$ &   1.       &  1.        &   1.       &  1.        &   1.       &  1.        \\
    $1$ &   0.557042 & -0.048552  &   0.557042 & -0.048552  &   0.557042 & -0.048552  \\
    $2$ &   1.70218  &  0.687291  &   1.83497  &  0.820079  &   1.83497  &  0.820079  \\
    $3$ &   2.43687  &  0.323257  &   2.83268  &  0.558242  &   3.01389  &  0.739452  \\
    \hline
    \end{tabular}
    \caption[tab:coeff-vf]{Perturbation series coefficients with $s=1$ for $\tilde{R}(q)$ ($a_l$) and $F(r)$ ($f_l$).
        Here $\as=\third$ for $a_3$ and $f_3$.}
    \label{tab:coeff-vf}
    \vspace*{2em}
    \begin{tabular}{|c|SS|SS|SS|}
    \hline
    & \multicolumn{2}{c|}{\MSbar} & \multicolumn{2}{c|}{geometric} & \multicolumn{2}{c|}{\cref{eq:alpha-new}, $n=2$} \\
    $l$ & {$v_l(1)$} & {$v_l(1)-V_l(1)$} & {$v_l(1)$} & {$v_l(1)-V_l(1)$} & {$v_l(1)$} & {\!$v_l(1)-V_l(1)$} \\
    \hline
    $0$ &   1.       &      0.206061     &   1.       &      0.182531     &   1.       &      0.177584     \\
    $1$ &   1.38384  &     -0.202668     &   1.38384  &     -0.249689     &   1.38384  &     -0.259574     \\
    $2$ &   5.46228  &      0.019479     &   5.59507  &     -0.009046     &   5.59507  &     -0.042959     \\
    $3$ &  26.6880   &      0.219262     &  27.3034   &      0.050179     &  27.4846   &      0.066468     \\
    \hline
    \end{tabular}
    \caption[tab:coeff]{Perturbation series coefficients with $s=1$ for $R(r)$ and $R_\text{RS}$ (with $V_l$ derived from $v_l$ as
        $R_l$ from $r_l$ in \cref{sec:MRS}).
        Here $\as=\third$ for $v_3$ and $v_l-V_l$.}
    \label{tab:coeff}
\end{table}
The scheme dependence in the two- and three-loop coefficients is about 10\%.
The (non)growth in~$l$ conforms with expectations: $a_l$ is perhaps growing slowly and $f_l$ is not growing yet.
(Recall, $p>1$ for $a_l$ and $p\ge9$ for $f_l$.) \Cref{tab:coeff} shows the first four $v_l$ in the same three schemes.
%
%
The growth is obvious.
\Cref{tab:coeff} also shows the subtracted coefficients $v_l(1)-V_l(1)$.
The cancellation is striking.

\begin{figure}
    \centering
    \includegraphics[width=\figwidth]{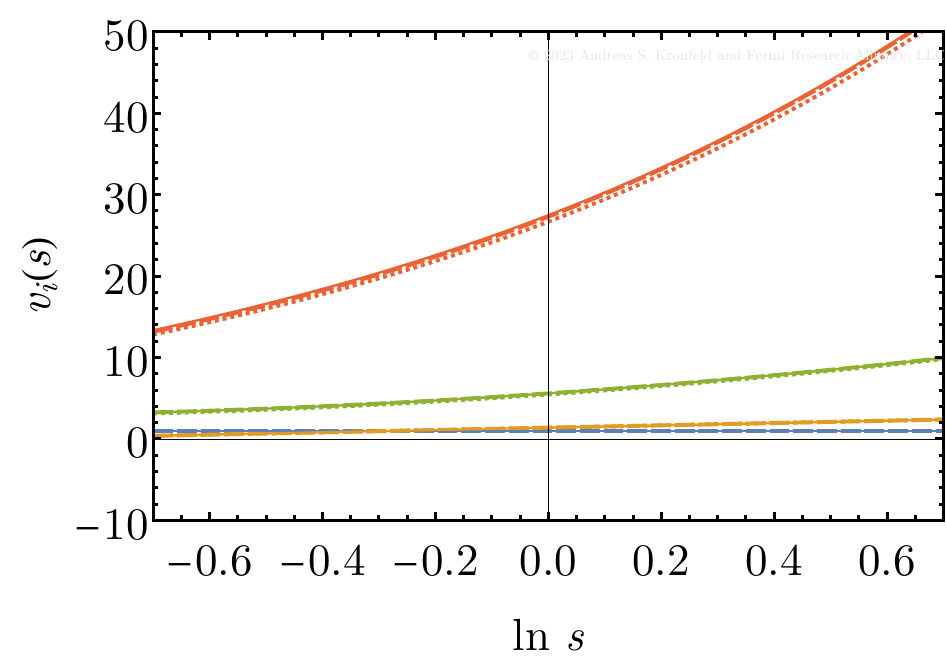} \hfill
    \includegraphics[width=\figwidth]{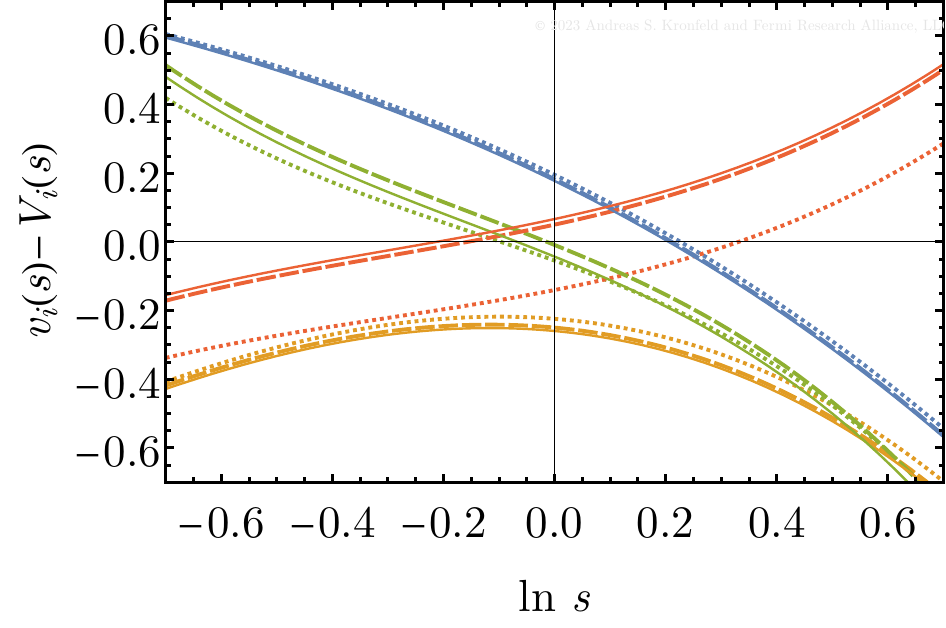}
    \caption[fig:vInset]{Scale dependence of $v_l(s)$ (left) and $v_l(s)-V_l(s)$ (right) vs.~$\ln s$.
        Note the difference in vertical scale.
        Blue, gold, green, and red correspond to $l=0$, $1$, $2$, and $3$, respectively.
        Dotted, dashed, and solid curves correspond to the \MSbar, geometric, and $\an{2}$ schemes, respectively.}
    \label{fig:vInset}
\end{figure}

The cancellation at $s=1$ is robust, as shown in \cref{fig:vInset} over an illustrative interval of~$\ln{s}$.
The range of $v_3(s)$ and even $v_2(s)$ dwarfs that of all $v_l(s)-V_l(s)$: $v_3(s)-V_3(s)$ ($v_2(s)-V_2(s)$) is 50--100 (5--10)
times smaller than $v_3(s)$ ($v_2(s)$).
Near $\ln s=0$, these two subtracted coefficients are unusually small.
Overall, the cancellation is best for $\ln{s}\approx\case{1}{4}$, where $|v_0-V_0|$ is especially small, while the others are of
typical size.

\begin{figure}
    \centering
    \includegraphics[width=\figwidth]{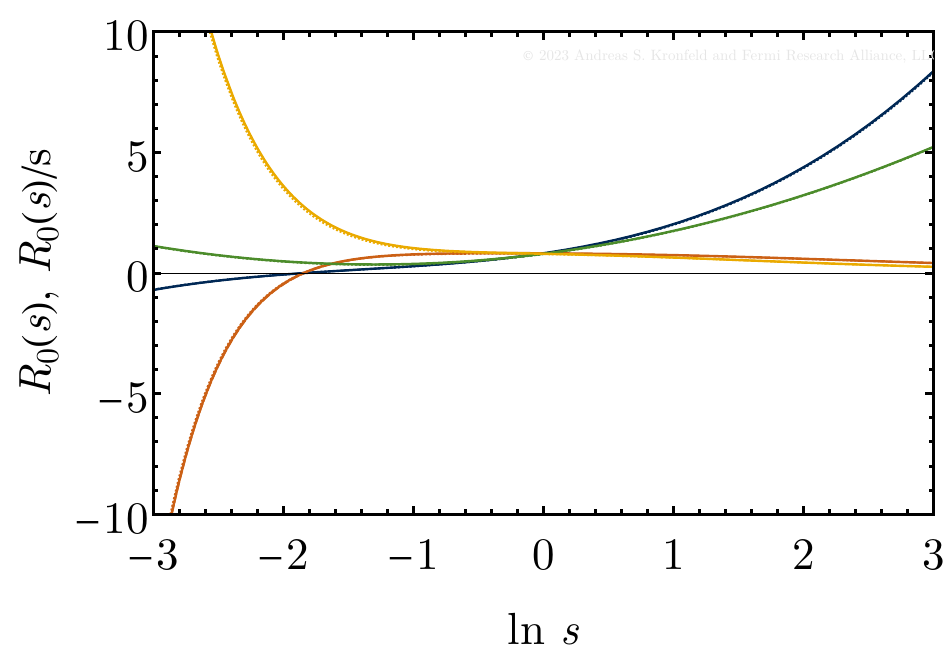} \hfill
    \includegraphics[width=\figwidth]{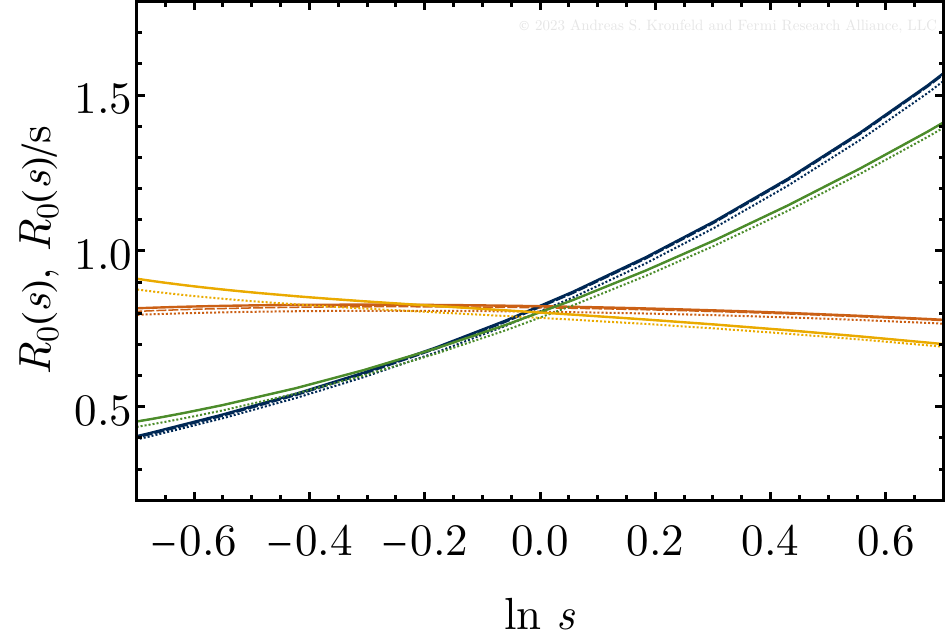}
    \caption[fig:V0]{Scale dependence of $R_0(s)$ and $R_0(s)/s$ over a very wide range (left) and a relevant range (right).
        Blue and orange (green and yellow) curves corresponds to $L=4$ ($L=3$) in \cref{eq:R0}.
        Blue and green (orange and yellow) curves corresponds to $R_0(s)$ ($R_0(s)/s$).
        Dotted, dashed, and solid curves correspond to the \MSbar, geometric, and $\an{2}$ schemes, respectively.}
    \label{fig:R0}
\end{figure}

Interestingly, as $\ln s$ is taken negative both factors in the first term $[v_l(s)-V_l(s)]\as(s/r)$ increase.
This behavior can be traced to the normalization factor $R_0(s)$, which is plotted in \cref{fig:R0} for the three schemes.
There is not much scheme dependence.
Curves for $L=4$ and $L=3$ in \cref{eq:R0} are shown.
They are close, or even very close, to each other for $|\ln s|\le\ln2$.
Sample numerical values are given in \cref{tab:R0}, again using both four and three terms in \cref{eq:R0}.
The shape of $R_0(s)$ follows from the positivity of the highest-power logarithmic term in \cref{eq:r} and the positivity of the
coefficients in \cref{eq:R0}.
Near $\ln s=-2$, the four-term $R_0(s)$ goes negative, which is a reflection of $v_3(s)$ being run to an absurd extreme while
omitting (unknown) higher orders.
Indeed, the three-term approximation to $R_0(s)$ turns up near $\ln s=-2$, which is a reflection of $v_2(s)$ being run to an absurd
extreme.
\begin{table}
    \centering
    \sisetup{table-format = 3.6, table-alignment-mode = format}
    \begin{tabular}{|c|SS|SS|SS|}
    \hline
    & \multicolumn{2}{c|}{\MSbar} & \multicolumn{2}{c|}{geometric} & \multicolumn{2}{c|}{\cref{eq:alpha-new}, $n=2$} \\
    $s$     &  {$L=4$} &  {$L=3$} &  {$L=4$} &  {$L=3$} &  {$L=4$} &  {$L=3$} \\
    \hline
    $\half$ & 0.386864 & 0.437281 & 0.403196 & 0.454397 & 0.397605 & 0.437281 \\
    $1$     & 0.793939 & 0.785114 & 0.817469 & 0.802230 & 0.801081 & 0.785114 \\
    $2$     & 1.52344  & 1.38707  & 1.55417  & 1.40419  & 1.52698  & 1.38707  \\
    \hline
    \end{tabular}
\caption[tab:coeff]{Normalization factor $R_0(s)$ of the $p=$ factorial growth in three schemes for $s\in\{\half, 1, 2\}$ at three
($L=4$) and two ($L=3$) loops.
Here $\as=\third$ for $L=4$.} \label{tab:R0}
\end{table}
\Cref{fig:R0} also shows $R_0(s)/s$, which multiplies the term absorbed into the power correction (cf., last sentence in
\cref{sec:MRS}).
It is nearly constant over a wide range, especially once $L=4$.

\begin{figure}
    \centering
    \includegraphics[width=\figwidth,trim={0 50 0 0},clip]{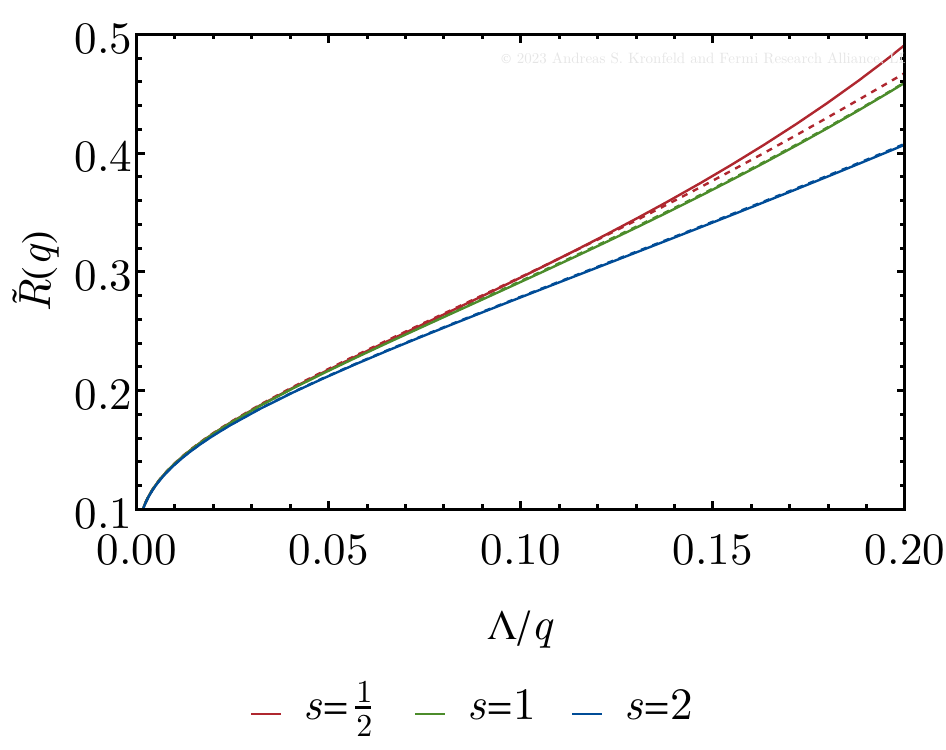} \hfill
    \includegraphics[width=\figwidth,trim={0 50 0 0},clip]{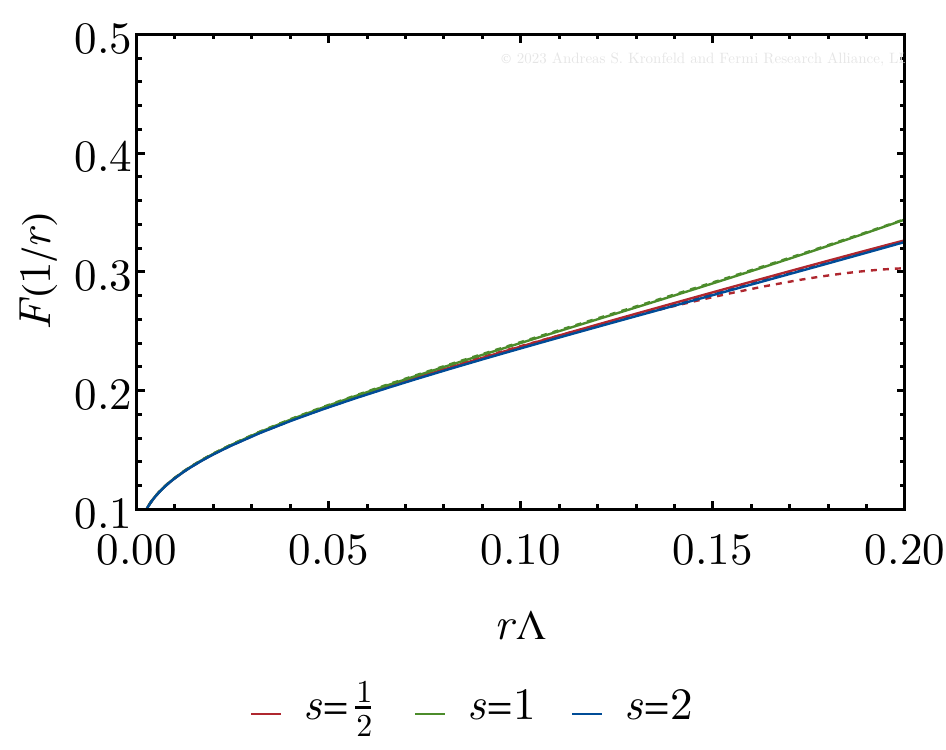} \\[6pt]
    \includegraphics[width=\figwidth,trim={0 50 0 0},clip]{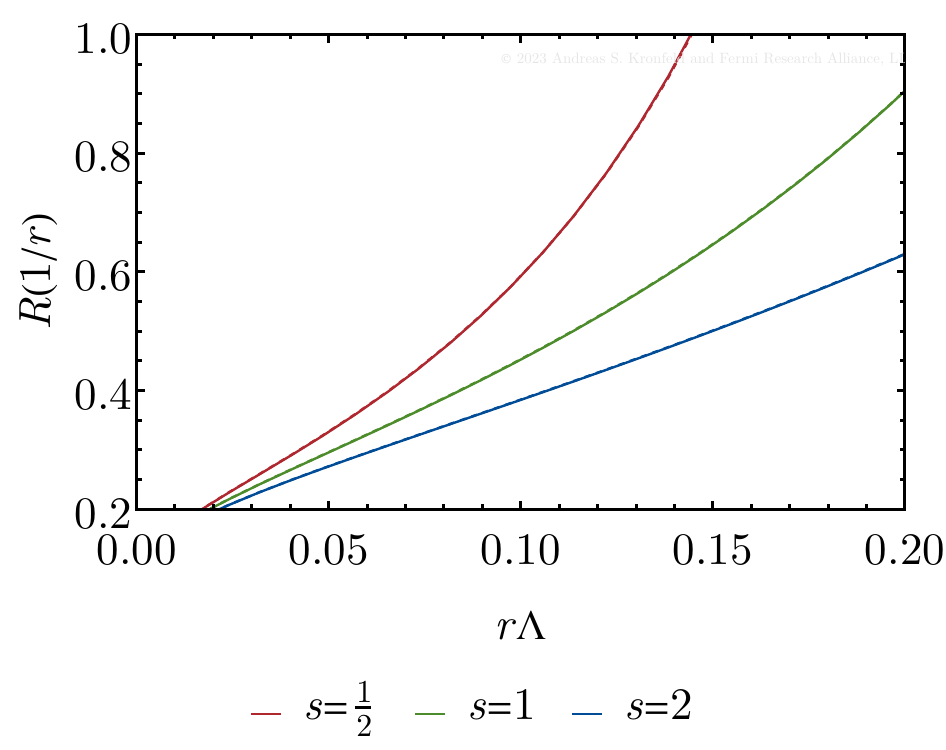} \hfill
    \includegraphics[width=\figwidth,trim={0 50 0 0},clip]{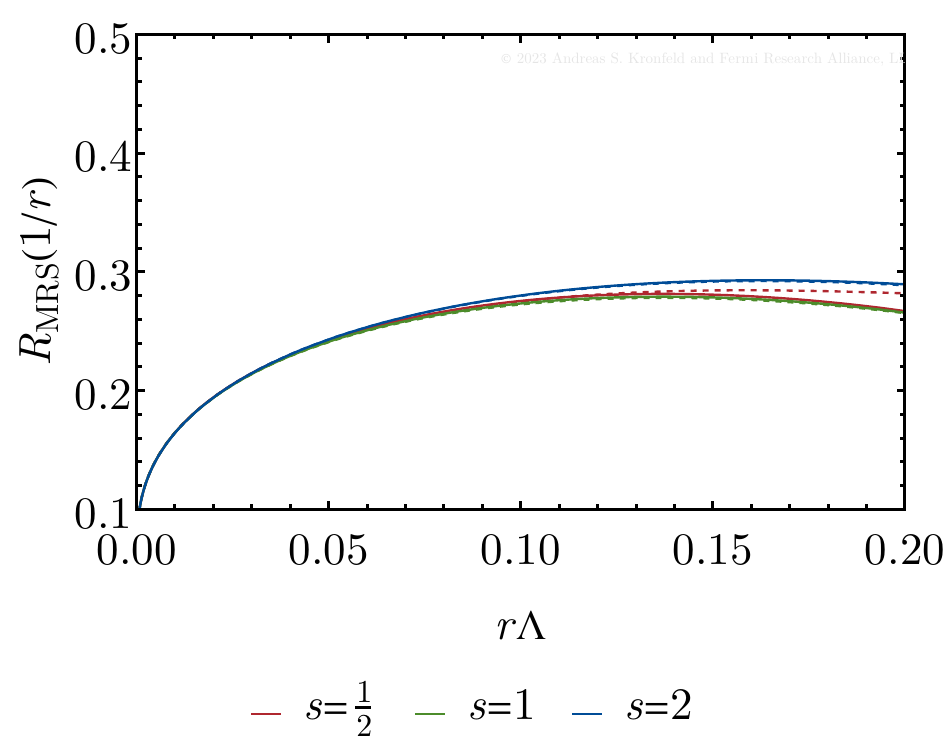}
    \caption[fig:PT]{Scale variation in the $\an{2}$ scheme up to and including the $\as^4$ term.
        Top: $\tilde{R}(q)$ and $F(1/r)$; neither suffers the $p=1$ renormalon.
        Bottom: $R(1/r)$ (with $p=1$ renormalon) and $R_\text{MRS}(1/r)$ (after MRS).
        Red, green, and blue curves correspond to $s=\half$, $s=1$, and $s=2$, respectively.
        Solid (dashed) curves correspond to a running (fixed) \as\ in the ultrasoft $\ln\as$.
        Note that the vertical scale for $R(1/r)$ is twice that of the other three plots.}
    \label{fig:PT}
\end{figure}

\begin{figure}
    \centering
    \includegraphics[width=\figwidth,trim={0 50 0 0},clip]{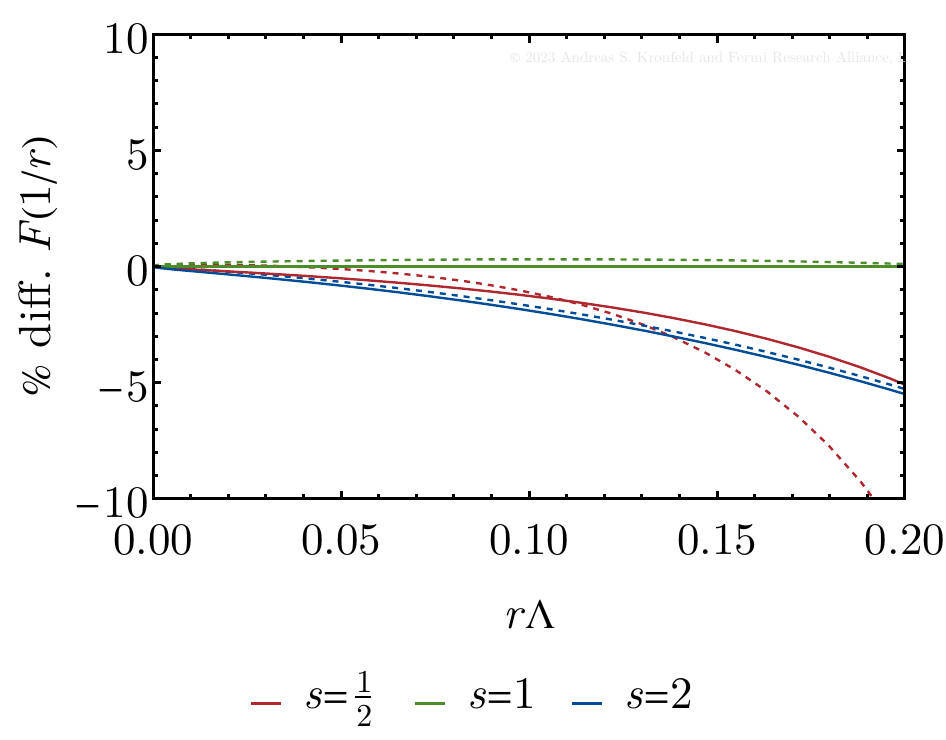} \hfill
    \includegraphics[width=\figwidth,trim={0 50 0 0},clip]{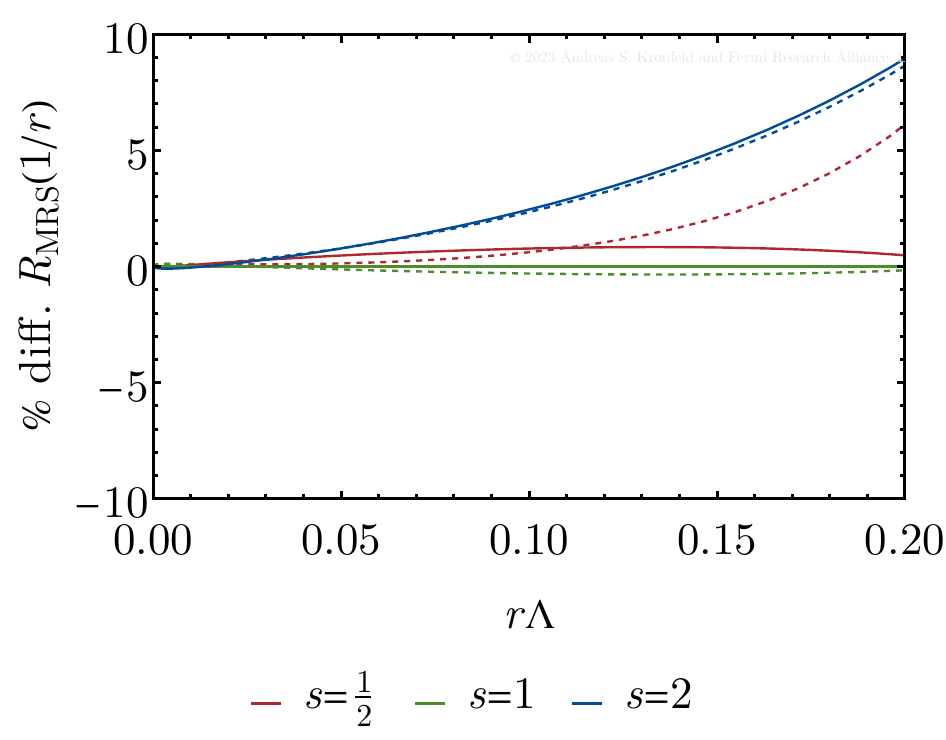}
    \caption[fig:fractional-scale-dependence]{Scale variation in the \an{2} scheme of the fractional difference of $F(1/r)$ (left)
        and $R_\text{MRS}(1/r)$ (right), with respect to $s=1$ with running ultrasoft \as.
        Curve and color code as in \cref{fig:PT}.}
    \label{fig:fractional-scale-dependence}
\end{figure}

The coefficients' variation with $s$ is set up to compensate that of $\as(sq)$ or $\as(s/r)$.
\Cref{fig:PT} shows how $\tilde{R}(q)$, $F(1/r)$, $R(1/r)$, and $R_\text{MRS}(1/r)$ depend on $\Lambda/q$ or $r\Lambda$ for
$s\in\{\half,1,2\}$.
(Plotted this way, the high-$q$, short-$r$ domain, where perturbation theory works best without any effort, is shrunk into a small
region.) %
The variation with $s$ is mild for $\tilde{R}(q)$, even milder for $F(1/r)$, and catastrophic for $R(1/r)$.
After MRS, however, the scale variation is as mild for $R_\text{MRS}(1/r)$ as for the renormalon-free~$F(1/r)$.
As shown in \cref{fig:fractional-scale-dependence}, the fractional difference of both remains a few percent for
$r\Lambda\lesssim0.1$ (with $s=1$ and running ultrasoft $\ln\as$ as the baseline).

\begin{figure}
    \centering
    \includegraphics[width=\figwidth,trim={0 50 0 0},clip]{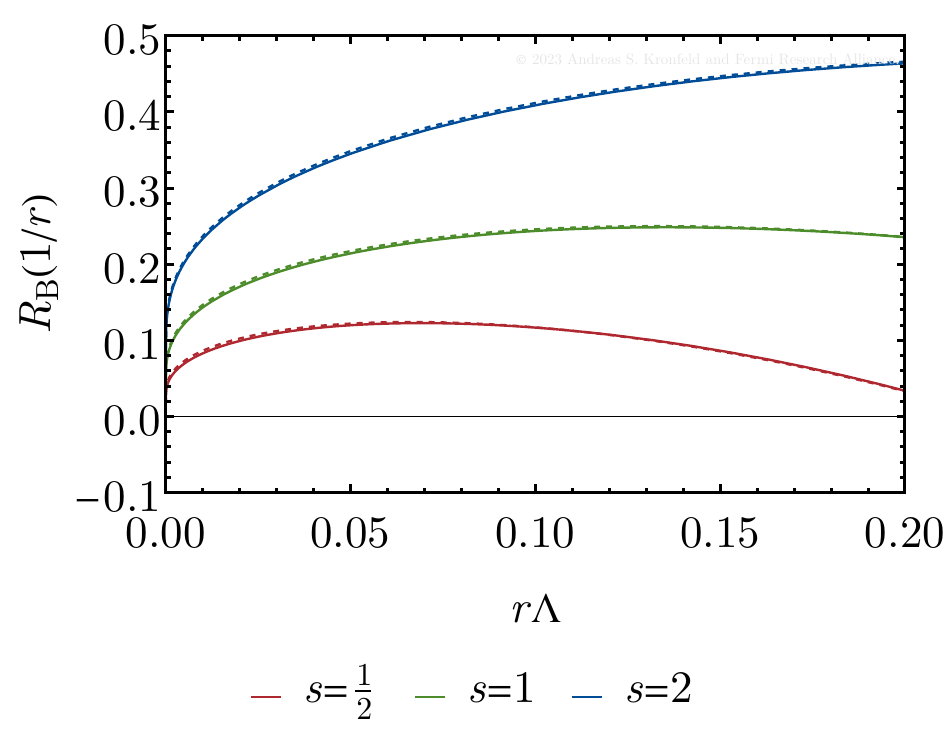} \hfill
    \includegraphics[width=\figwidth,trim={0 50 0 0},clip]{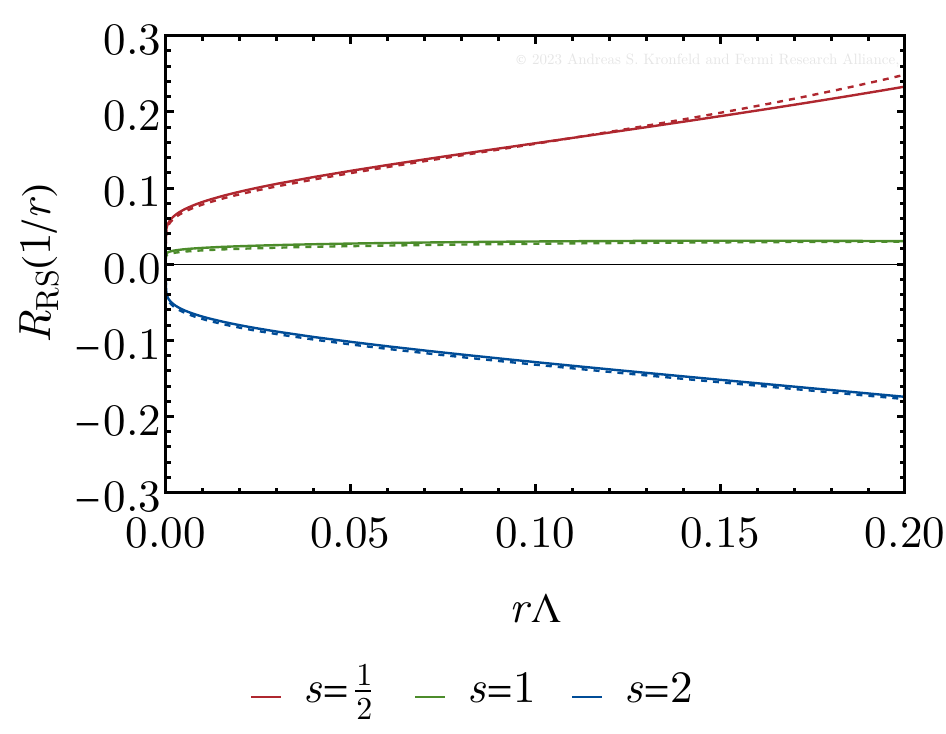}
    \caption[fig:cancel]{Scale variation in the \an{2} scheme of the Borel sum $R_\text{B}(1/r)$~(left) and the
        $L=4$-sub\-tracted series $R_\text{RS}(1/r)$~(right).
        Curve and color code as in \cref{fig:PT}.}
    \label{fig:cancel}
\end{figure}

The mild variation with $s$ is a pleasant outcome given the $s$ dependence of the subtracted coefficients (cf., \cref{fig:vInset}).
\Cref{fig:cancel} shows the variation with $s$ as a function of~$r$ of the Borel sum $R_\text{B}(1/r)$~(left, \cref{eq:RJ}) and the
subtracted series $R_\text{RS}(1/r)$ for $L=4$ (right).
Both are quite sensitive to~$s$, but their sum (bottom right of \cref{fig:PT}) is not.

\begin{figure}
    \centering
    \captionbox{Borel sum $R_\text{B}(1/r)$ (dotted curves) accumulating successively each term $(v_l-V_l)\as^{l+1}$ (dashed
        curves with longer dashes for larger~$l$) in the three schemes.
        Solid curves for the full $R_\text{MRS}$.
        Color code as in \cref{fig:PT}.\label{fig:order}}[0.49\textwidth]{%
    \includegraphics[width=0.49\textwidth,trim={10 50 0 0},clip]{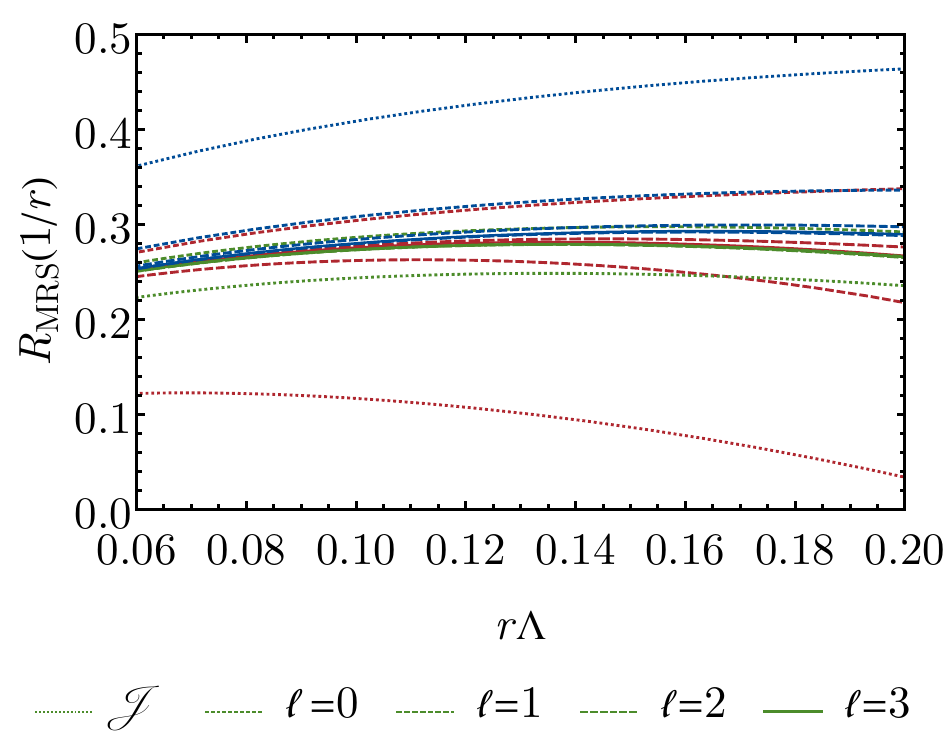}} \hfill
    \captionbox{Same as \cref{fig:PT} (bottom right) but with a band stemming from the uncertainty in $R_0$ (taken equal to
        the last term in \cref{eq:R0}) and an expanded vertical scale.
        Curve and color code as in \cref{fig:PT}.\label{fig:V0err}}[0.49\textwidth]{%
    \includegraphics[width=0.49\textwidth,trim={10 50 0 0},clip]{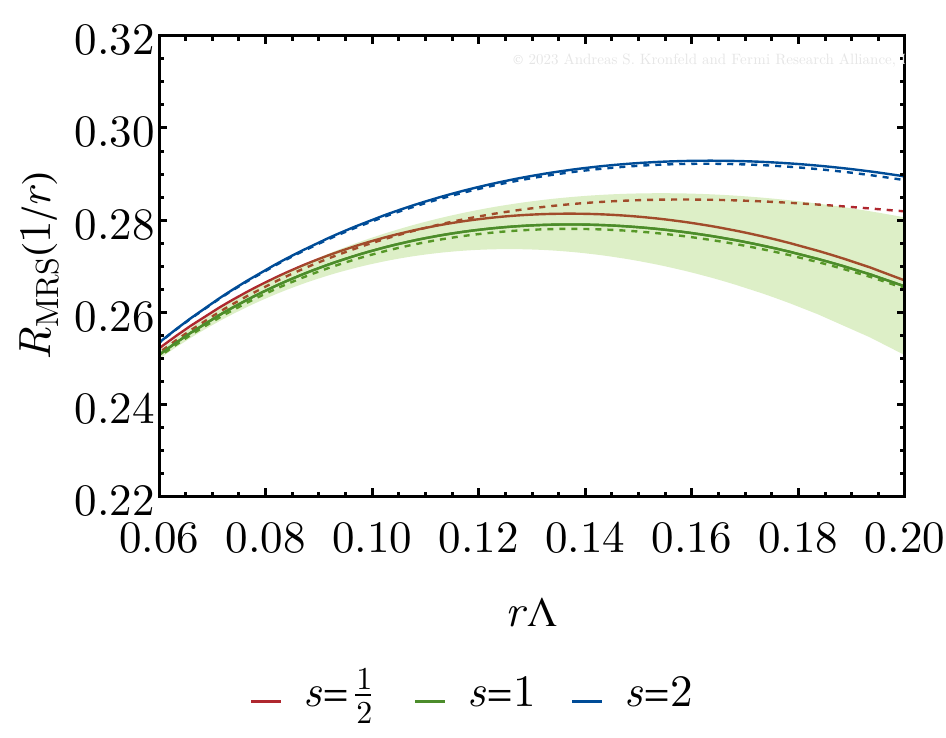}}
\end{figure}

The first two orders suffice to lift the $s$ dependence, as shown in \cref{fig:order}.
Here, $R_\text{B}(1/r)$ is shown (dotted curve) and each term $(v_l-V_l)\as^{l+1}$, $l=0,1,2,3$, in $R_\text{RS}(1/r)$ is
accumulated (dashed curves with longer dashes as the order increases) until the total $L=4$ (solid) result $R_\text{MRS}(1/r)$ is
reached.
Adding the tree-level term $(v_0-V_0)\as$ to the Borel sum overshoots the full (solid) result, but adding the one-loop term
$(v_1-V_1)\as^2$ yields a curve almost indistinguishable from $R_\text{MRS}(1/r)$.
Indeed, it is hard to distinguish the longer-dashed curves from the solid ones, underscoring that the two-loop term $(v_2-V_2)\as^3$
makes a small change while the three-loop term $(v_3-V_3)\as^4$ makes hardly any change.
As with the pole mass~\cite{Brambilla:2017hcq}, MRS perturbation theory converges (in the practical sense) quickly.

Let us return to the term-by-term change in $R_0$ (cf., \cref{fig:R0} and \cref{tab:R0}).
The highest-order term in \cref{eq:R0} can be used to estimate the uncertainty in $R_0$ from omitting even higher
orders~\cite{Brambilla:2017hcq}.
In the case at hand, the term with $f_3$ yields the estimate, which is around $10\%$ or less (cf., \cref{tab:R0}).
\Cref{fig:V0err} (right) overlays the resulting $s=1$ uncertainty band on $R_\text{MRS}(1/r)$ on the curves (at various $s$) of
\cref{fig:PT} (bottom right).
The uncertainty propagated to $R_\text{MRS}$ is smaller than $10\%$ because changes in $R_0$ push $R_\text{B}(1/r)$ and
$R_\text{RS}(1/r)$ in opposite directions.
Indeed, the uncertainty in $R_\text{MRS}$ stemming from $R_0$ is smaller than the difference between the $s=1$ and the $s=\half$
and~2 curves.
Note, however, that the $R_0$-uncertainty, as defined here, is smaller at $s=1$ than at $s=\half$ and~2 (cf., \cref{fig:R0} and
\cref{tab:R0}).
The uncertainty bands of these other choices (not shown) cover all three.

\section{Summary and outlook}
\label{sec:outlook}

The initial aim of this work was to study and extend the discussion of factorial growth and renormalons started in
refs.~\cite{Komijani:2017vep,Brambilla:2017hcq}.
I found, however, that the perspective, derivation, and interpretation could be simplified: a straightforward analysis extracts
information from the renormalization-group constraints on the series coefficients.
The only other ingredient is the knowledge (or assumption) of the powers $\{p_i\}$ of the power-suppressed corrections to the
perturbative series of a physical observable.
A by-product of adding this information to the series is to subtract the leading factorial growth (aka ``renormalon effects'') from
the first few series coefficients.
Remarkably, the factorial growth is not just a large-order phenomenon: it starts at low orders.
How it comes to dominate the coefficients depends of the power of the power correction.
(The lower the power, the more powerful the factorial!)

The worked example of the static energy (\cref{sec:E0}) seems successful in removing a power correction of order $r\Lambda$ from
(a~dimensionless version of the) static energy, $-rE_0/C_F$.
The conventional choice of $\mu=1/r$ (in the \MSbar\ scheme) seems near an optimum: perturbation theory converges with the MRS
treatment as well as it does for the static force, which is thought to suffer corrections only of high power.
Varying $\mu$ by a factor of two rearranges contributions between the tree and one-loop fixed order contributions, on the one hand,
and (a specific definition of) the Borel sum of the factorial growth, on the other.
At \emph{very} short distances, $r\Lambda\lesssim\case{1}{16}$, the total result (using all information through order~$\as^4$) does
not vary over a wide range of~$\mu$.
It will therefore be interesting to fit to lattice-QCD data (e.g., that of ref.~\cite{Brambilla:2022het}) and compare with other
approaches to taming the series.
(Some other approaches are described in
refs.~\cite{Takaura:2018vcy,Bazavov:2019qoo,Ayala:2020odx,Ananthanarayan:2020umo,Brambilla:2021wqs,Sumino:2020mxk} and earlier work
cited there.)

Some practical issues remain before applying the MRS procedure to, say, an \as\ determinations.
The MRS method, like standard perturbation theory, does not say what scale $s$ to choose: starting in the \MSbar\ scheme with $s=1$
and varying by a factor of~2 is conventional.
When the factorial growth of coefficients matters, i.e., when MRS has something to offer, the $\ln^ns$ contributions associated with
scale setting cannot tame the coefficients.
Scale setting in light of MRS may warrant a closer look.
Another issue is that in many applications, some of the quarks cannot be taken massless.
A nonzero quark mass in a loop alters the loop's growth, removing the factorials from the infrared.
It is probably best to add massive quark-loop effects at fixed order and not to use the massless result as a stand-in for the
massive one~\cite{Hoang:2000fm}.
Last, when anomalous dimensions are an important feature for more than one round of MRS, the method (as presented in
\cref{sec:anom}) remains to cumbersome to be appealing.
It may suffice to neglect the anomalous dimensions, but only practical experience will tell.

\appendix

\section{Modified Borel summation}
\label{app:MBS}

Alternatives to the standard Borel resummation are possible~\cite{Brown:1992pk}, and a natural variant is pursued here, leading to
the same endpoint.
Start with \cref{eq:RRS}:
\begin{equation}
    R_\text{B}^{(p)} \equiv \sum_{l=0}^\infty R_l^{(p)} \alpha^{l+1} = R_0^{(p)} \alpha \sum_{l=0}^\infty
        \left(\frac{2\beta_0\alpha}{p}\right)^l \frac{\Gamma(l+1 + pb)}{\Gamma(1 + pb)} .
    \label{eq:R-appendix}
\end{equation}
The $l$-dependent $\Gamma$ function can be expressed as $\Gamma(l+1 + pb) = \int_0^\infty t^{l+pb}\e^{-t}\dd{t}$.
Swapping the order of summation and integration
\begin{equation}
    R_\text{B}^{(p)} = \frac{R_0^{(p)} \alpha}{\Gamma(1 + pb)} \int_0^\infty \sum_{l=0}^\infty
        \left(\frac{2\beta_0\alpha t}{p}\right)^l t^{pb}\e^{-t}\dd{t}
        = \frac{R_0^{(p)} \alpha}{\Gamma(1 + pb)} \int_0^\infty \frac{t^{pb}\e^{-t}}{1-2\beta_0\alpha t/p}\dd{t},
    \label{eq:R-Borel-mod}
\end{equation}
which only has a simple pole instead of a branch point.
After integrating
\begin{equation}
    R_\text{B}^{(p)} = R_0^{(p)} \frac{p}{2\beta_0} \mathcal{J}(pb,p/2\beta_0\alpha)
        - R_0^{(p)} \e^{\pm\iunit pb\pi} \frac{p^{1+pb}}{2^{1+pb}\beta_0} \Gamma(-pb)
            \left[\frac{\e^{-1/2\beta_0\alpha}}{(\beta_0\alpha)^b} \right]^p ,
    \label{eq:R-Borel-mod-sum}
\end{equation}
where the factor $\e^{\pm\iunit pb\pi}$ in the second term corresponds to passing the contour below or above the pole.
As before, the second term can be absorbed into the power correction, and the first --- the principal part --- is taken to define
$R_\text{B}^{(p)}$, the same as \cref{eq:RB}.

\acknowledgments

This work is supported in part by the Technical University of Munich, Institute for Advanced Study, funded by the German Excellence
Initiative.
Fermilab is managed by Fermi Research Alliance, LLC, under Contract No.\ DE-AC02-07CH11359 with the U.S.\ Department of Energy.


\bibliographystyle{JHEP}
\bibliography{j}

\end{document}